\documentclass[lettersize,journal]{IEEEtran} 
\usepackage{amsmath,amssymb,amsfonts,amsthm}
\usepackage{hyperref}
\usepackage{xcolor}
\usepackage{algorithmic}
\usepackage{graphicx}
\usepackage{textcomp}
\usepackage{amsmath,amsfonts}
\usepackage{algorithmic}
\usepackage{algorithm}
\usepackage{array}
\usepackage{cite}
\usepackage{textcomp}
\usepackage{url}
\usepackage{verbatim}
\usepackage{bm} 
\usepackage{booktabs}
\usepackage{enumitem}
\usepackage{siunitx}
\hypersetup{colorlinks=true, linkcolor=blue, citecolor=blue}
\usepackage[capitalize,noabbrev]{cleveref}
\usepackage{subcaption}
\usepackage{tikz, pgfplots}
\usetikzlibrary{plotmarks,spy,angles,quotes}
\pgfplotsset{compat=newest}
\usepackage{caption}
\usepackage{cuted}
\usepackage{nccmath}

\definecolor{blue1}{RGB}{0,114,189}
\definecolor{orange1}{RGB}{217,83,25}
\definecolor{yellow1}{RGB}{237,177,32}
\definecolor{purple1}{RGB}{126,47,142}
\definecolor{green1}{RGB}{119,172,48}
\definecolor{red1}{RGB}{162,20,47}

\DeclareMathOperator{\diag}{diag}

\DeclareMathOperator{\sinc}{sinc}
\newcommand{\T}{{\sf T}}
\renewcommand{\H}{{\sf H}}

\newcommand{\J}{{\mathbf{J}}}

\newcommand{\U}{{\mathbf{U}}}

\newcommand{\bPhi}{{\boldsymbol{\Phi}}}
\definecolor{RED}{rgb}{0.7,0,0}
\definecolor{BLUE}{rgb}{0,0,0.69}
\definecolor{GREEN}{rgb}{0,0.6,0}
\definecolor{PURPLE}{rgb}{0.69,0,0.8}

\newtheorem{Theorem}{Theorem}
\newtheorem{Corollary}{Corollary}

\let\texdisplaystyle\displaystyle
\def\displaytotextstyle{\textstyle\let\displaystyle\texdisplaystyle}

\newenvironment{talign*}
 {\let\displaystyle\displaytotextstyle\csname align*\endcsname}
 {\endalign}

\newcommand{\mb}[1]{\mathbf{#1}}

\def\BibTeX{{\rm B\kern-.05em{\sc i\kern-.025em b}\kern-.08em
    T\kern-.1667em\lower.7ex\hbox{E}\kern-.125emX}}
\usepackage{balance}
\begin{document}
\title{Metasurface-Enabled Extremely Large-Scale Antenna Systems: Transceiver Architecture, Physical Modeling, and Channel Estimation}

\author{Zhengyu~Wang, ~\IEEEmembership{Student Member,~IEEE}, 
Gui~Zhou, ~\IEEEmembership{Member,~IEEE},
Tiebin~Mi, ~\IEEEmembership{Member,~IEEE},
Rujing~Xiong, ~\IEEEmembership{Member,~IEEE},
Jianan~Zhang, ~\IEEEmembership{Member,~IEEE},
Robert C.~Qiu,~\IEEEmembership{Fellow,~IEEE}
\thanks{Z.~Wang, G.~Zhou, T.~Mi, J.~Zhang and R.~C.~Qiu are with the School of Electronic Information and Communications (EIC), Huazhong University of Science and Technology (HUST), Wuhan 430074, China. (e-mail:\{wangzhengyu, gui\_zhou, mitiebin, zhangjn, caiming\}@hust.edu.cn.). R.~Xiong is with the School of Science and Engineering, The Chinese University of Hong Kong, Shenzhen, Shenzhen 518172, China (e-mail: rujingxiong@cuhk.edu.cn).
\textit{Corresponding author: Tiebin~Mi.} 
% (email: \href{mailto:zhenyu_liao@hust.edu.cn}{zhenyu\_liao@hust.edu.cn}).
}
\thanks{This work is supported by the National Natural Science Foundation of China under Grant 12141107, the Key Research and Development Program of Wuhan under Grant 2024050702030100, and the Interdisciplinary Research Program of HUST (2023JCYJ012).}
% \thanks{The code in this paper is available online at https://github.com/zhengyuwang0/GESPRIT.}
}

\maketitle

\begin{abstract}
Extremely large-scale antenna arrays (ELAAs) have emerged as a pivotal technology for addressing the unprecedented performance demands of next-generation wireless communication systems. 
To enhance their practicality, we propose metasurface-enabled extremely large-scale antenna (MELA) systems--novel transceiver architectures that replace the phase-shifter-antenna front end with a reconfigurable transmissive metasurface, enabling a few active feeds to wirelessly excite a large passive aperture.
This architecture seliminates the need for bulky switch matrice and costly phase-shifter networks typically required in conventional solutions. 
Physically grounded models are developed to characterize electromagnetic field propagation through individual transmissive unit cells, capturing the fundamental physics of wave transformation and transmission. 
Additionally, distance-dependent approximate models are introduced, exhibiting structural properties conducive to efficient parameter estimation and signal processing. 
Based on the channel model, a two-stage channel estimation framework is proposed for the scenarios comprising users in the hybrid near- and far-fields. 
In the first stage, a dictionary-driven beamspace filtering technique enables rapid angular-domain scanning. In the refinement stage, the rotational symmetry of subarrays is exploited to design super-resolution estimators that jointly recover angular and range parameters. An analytical expression for the half-power beamwidth of MELA is derived, revealing its near-optimal spatial resolution relative to conventional ELAA architectures. 
Numerical experiments further validate the high-resolution of the proposed channel estimation algorithm and the fidelity of the electromagnetic model, positioning the MELA architecture as a highly competitive and forward-looking solution for practical ELAA deployment.

\end{abstract}


\section{Introduction}
\label{sec:intro}

The deployment of multiple antennas has long been a cornerstone in the evolution of wireless communications, enabling foundational technologies such as MIMO systems and spatial beamforming \cite{mendez2016hybrid,10379539}.
% {\color{red}
As the vision for sixth-generation (6G) and next-generation wireless networks continues to evolve, extremely large-scale antenna arrays (ELAAs), also known as extremely large-scale multiple-input multiple-output (XL-MIMO), have emerged as a key enabler to meet the unprecedented demands for spectral efficiency \cite{wang2023extremely, yu2024patternedbeamtrainingnovel}, spatial resolution \cite{9860979}, and massive user connectivity \cite{10734395}. 
% }
By substantially increasing both the aperture size and the number of radiating elements, ELAAs introduce new spatial degrees of freedom that fundamentally transform the wireless interface.

\begin{figure}[t]
    \centering
    \includegraphics[width=\linewidth]{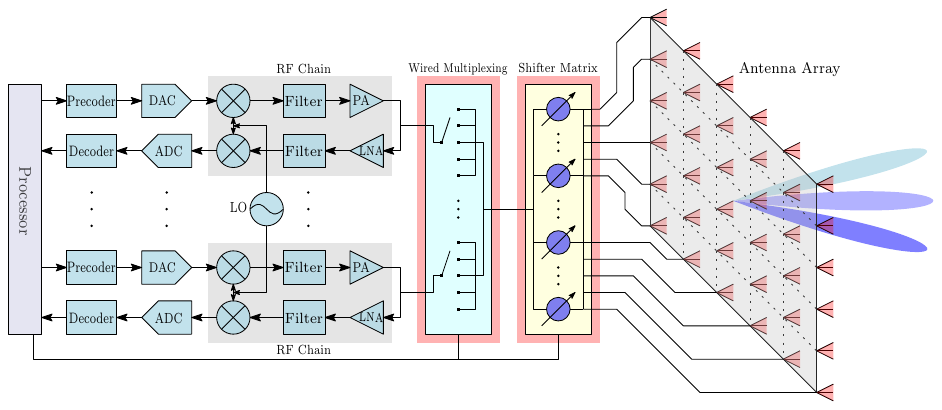}
    \caption{Illustration of the conventional ELAA architecture.
    % , where high-overhead switches and phase shifters dynamically manage the connections between a limited number of RF chains and the full antenna array.
    }\label{fig:elaa}
    \vspace{-6pt}
\end{figure}

One of the most critical and persistent challenges in the practical realization of ELAAs is efficiently interfacing a massive number of antenna elements with a limited set of radio frequency (RF) chains \cite{xiao2017millimeter}. 
While one-to-one RF-to-antenna mapping remains marginally feasible in conventional massive MIMO systems, it becomes impractical at ELAA scales due to prohibitive costs, power and implementation complexity \cite{mendez2016hybrid}. 
Moreover, interconnect complexity and phase-shifter count scale with the product of RF chains and array size.
As a compromise, partially connected architectures have been proposed, wherein a reduced number of RF chains are dynamically linked to the antenna array via networks of switches \cite{han2023toward,10845800}, as illustrated in Fig.~\ref{fig:elaa}. 
Although these multiplexing schemes offer a more cost- and energy-efficient alternative to full connectivity, they still involve substantial hardware complexity and bulky circuitry, ultimately limiting the scalability of the system.
These challenges may partly explain the current lack of large-scale ELAA hardware prototypes.

Recent advances in metasurface technologies \cite{9365009} have introduced a new paradigm for low-cost and flexible phase control, offering a promising solution for simplifying transceiver architectures via over-the-air replacement of conventional switch and phase-shifter networks.
This paper proposes metasurface-enabled extremely large-scale antenna (MELA) systems—transceiver architectures for scalable, hardware-efficient ELAAs.
Unlike traditional wired implementations, MELA systems exploit the wireless channel for transceiver-to-antenna coupling and delegate phase control to the metasurface, as illustrated in Fig.~\ref{fig:architecture}, eliminating the need for bulky switch matrices and phase-shifter networks.
To realize wireless coupling, auxiliary feed antennas are deployed at the RF end to illuminate the metasurface.
This architecture enables seamless expansion of both RF chains and antenna elements without requiring substantial modifications to the underlying design.

\vspace{-5pt}
\subsection{Contributions}
The main contributions of this work are summarized as follows:

\begin{enumerate}[label=\textbf{\arabic*)}, leftmargin=*, nosep]
\item \textbf{Transmissive metasurface-enabled over-the-air architecture:} 
We propose a novel transceiver architecture for ELAAs that leverages reconfigurable transmissive metasurfaces. 
Unlike conventional ELAA systems that rely on dense networks of switches and phase shifters, the proposed MELA architecture utilizes the wireless channel for transceiver-to-antenna coupling and delegates phase control to the metasurface. 
This over-the-air design substantially reduces hardware complexity and signal routing overhead, while offering greater architectural flexibility and scalability for modular expansion.

\item \textbf{Physically interpretable and mathematically tractable channel modeling for MELA:} 
We propose a theoretical framework to accurately model the end-to-end wireless link--from the source to the transmissive metasurface and finally to the receiver aperture--as a cascaded interaction of electromagnetic fields. 
We obtain closed-form expressions for the electric field and establish a matrix-form channel representation. 
Based on the placement distances between the metasurface and receivers, we further derive distance-specific approximate channel models that not only offer accurate characterization but also exhibit desirable structural properties for efficient analysis and processing.

\item \textbf{Hybrid-field and gridless channel estimation for MELA:}
We propose a two-stage uplink channel estimation framework, supporting both near- and far-field signal components.
In the coarse stage, a dictionary-based beamspace filtering technique enables fast angular-domain scanning through dynamic phase configuration of the metasurface.
In the refinement stage, the subarray rotational symmetry is exploited to construct super-resolution estimators that jointly recover angle and distance parameters.
Furthermore, we derive the half-power beamwidth (HPBW) expression for the MELA system and show that its spatial resolution closely approximates that of traditional ELAA architectures under practical conditions, thereby validating its effectiveness for high-precision sensing and channel acquisition.

\end{enumerate}

\vspace{-3pt}
\subsection{Prior Works}
\begin{enumerate}[label=\textbf{\arabic*)}, leftmargin=*, nosep]
\item \textit{Hardware Architecture:} 
% With the rapid progress of electromagnetic metamaterials, reconfigurable intelligent surfaces have emerged as a promising technology.
% These inherently passive surfaces typically operate without dedicated RF chains and manipulate incident waves by simply reflecting \cite{xiong2023ris} or transmitting \cite{tang2023transmissive} the signals.
% Embedding passive metasurfaces within transceiver architectures has emerged as a compelling research direction.
The concept of reconfigurable transmitarrays was first introduced in \cite{1143726}, where the authors proposed a sandwich-structured planar three-dimensional lens. This architecture comprises a receiving antenna array phase-coupled to a transmitting array via tunable phase shifters, jointly illuminated by one or multiple focal sources \cite{6352828,7502105}.
More recently, Cui \textit{et al.} \cite{wang2024hybrid} proposed a hybrid metasurface lens architecture paired with a programmable uniform linear feed array along the focal line.
However, all these designs rely on a horn antenna placed on the focal line as the excitation source, a configuration that is further extended in this work. 
Another promising architecture is the dynamic metasurface antenna (DMA) \cite{shlezinger2021dynamic}, which embeds a dense array of metasurface elements along waveguides; the propagating signal is coupled and radiated to perform analog-domain weighting and combining \cite{jabbar202460}.
This approach drastically reduces the number of digital ports relative to radiating elements, and thus lowers RF chain count and interconnect complexity.
Despite its cost-effectiveness, the DMA suffers from attenuation of the guided wave along the feeding waveguides \cite{boyarsky2021electronically}, which constrains aperture scaling and becomes even more severe in two-dimensional feed networks.

\begin{figure}[t]
    \centering
    \includegraphics[width=0.95\linewidth]{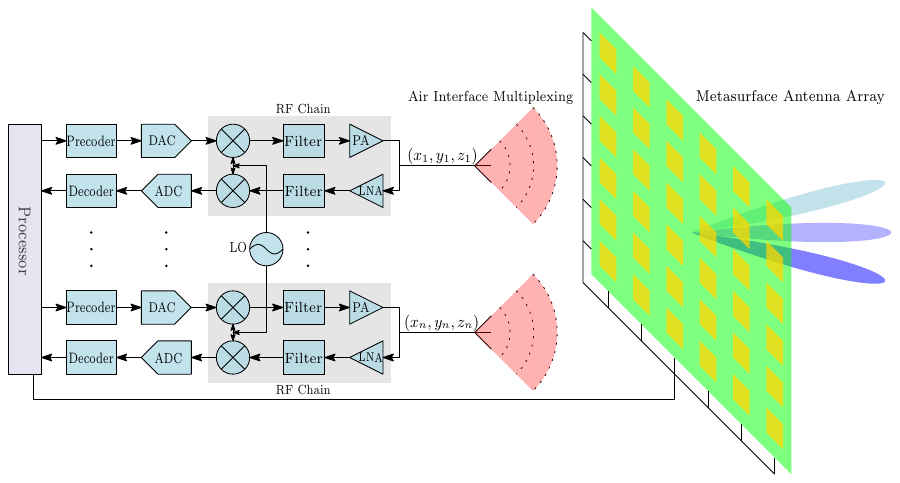}
    \vspace{-6pt}
    \caption{Illustration of the MELA architecture.
    % , which employs over-the-air transceiver-to-antenna coupling and delegates phase control to the reconfigurable metasurface.
    }    
\label{fig:architecture}
\vspace{-6pt}
\end{figure}
\item \textit{Theoretical Modeling:}
In theoretical research, the structure of a reconfigurable metasurface integrated with a feed antenna was employed in \cite{li2021beamforming,li2023robust,li2024transmissive2} for signal modulation and downlink wireless transmission.
However, these studies share a critical limitation: the wireless channel between the feed antenna and the metasurface is often ignored, and the transmitter is modeled as a unified black-box entity.
Subsequently, \cite{li2024transmissive} established transmission schemes for both downlink and uplink communications under this architecture.
For uplink systems, \cite{li2023toward} proposed a joint optimization framework that simultaneously adjusts power allocation and the phase-shift coefficients of the transmissive metasurface to maximize the system sum-rate, and \cite{demir2024user} investigated this transceiver architecture in cell-free massive MIMO uplink scenarios.
These studies commonly assume that the distance between the transmissive metasurface and the receiver lies within the Rayleigh zone, where the metasurface-receiver link can be characterized by a line-of-sight near-field propagation model \cite{li2024transmissive, li2023toward, demir2024user}.
However, in the sub-aperture near-field region, the finite size of the antenna elements is no longer negligible, as shown experimentally in \cite{li2024transmissive}.
Prior works model each antenna element as a point source, which makes conventional channel models insufficient for precisely characterizing the behavior of reconfigurable transmitarrays in this regime.
\item \textit{Channel Estimation:}
As the Fresnel region significantly expands in ELAA systems, signal sources may reside in either the near-field or far-field regime \cite{han2024cross}.
When both near-field and far-field signals coexist, near-field steering vectors introduce energy spreading along the distance dimension for far-field signals, while far-field steering vectors cause angular-domain leakage for near-field signals \cite{wu2024near}.
These power diffusion effects pose major challenges to conventional channel estimation methods, which typically assume a single-field scenario \cite{10620366}.
Moreover, due to the inherent non-orthogonality of near-field steering vectors \cite{cui2022channel}, conventional sparse recovery algorithms such as orthogonal matching pursuit (OMP) in the polar domain \cite{wei2021channel} become ineffective.
To address this issue, several recent studies have proposed hybrid-field sparse channel representations \cite{wang2023compressive,xi2023gridless}.
% For instance, \cite{wang2023compressive} decouples the angle- and distance-related components in the steering vectors and formulates a block-sparse compressed sensing problem.
% In another line of work, \cite{xi2023gridless} leverages hybrid-field sparsity in the fractional Fourier domain and applies iterative refinement to mitigate off-grid errors.
Beyond compressed sensing, \cite{ramezani2024efficient} exploits array geometry symmetry and employs a modified MUSIC algorithm to decouple angle and distance estimation \cite{he2011efficient}.
However, the iterative scheme in \cite{xi2023gridless} incurs substantial computational overhead, while the spectral search methods in \cite{ramezani2024efficient, he2011efficient} suffer from large search spaces and high complexity.
Additionally, sparse-based hybrid channel estimation approaches face fundamental limitations, including grid mismatch errors that constrain estimation accuracy and the use of excessively dense grids that increase dictionary size.
Despite these advances, high-dimensional channel estimation in ELAA systems remains a critical and open research challenge.
\end{enumerate}

% The proposed MELA system offers several key advantages over conventional ELAA architectures. First, the air-interface-based multiplexing scheme leverages the spatial wireless channel as the medium for transceiver-to-antenna coupling, thereby eliminating the need for complex switch matrices and extensive cabling. Second, phase shifting is implemented directly by the metasurface, removing the requirement for dedicated phase shifter networks. This approach significantly reduces hardware complexity and enhances the scalability and integration flexibility of the overall system. 
% For example, the proposed architecture supports seamless scaling of both RF chains and antenna elements without necessitating substantial modifications to the underlying system design.

The remainder of this paper is organized as follows.
In \Cref{sec:system model}, the MELA architecture and corresponding channel models are introduced. 
% Additionally, we examine two special cases and derive their corresponding approximate channel models.
In \Cref{sec:channel es}, we present a channel estimation scheme specifically designed for the MELA architecture.
In \Cref{sec:sp_resolution}, we derive an expression for HPBW of the MELA and compare its spatial resolution with that of conventional ELAA architectures.
% Simulation results validating our theoretical channel model and estimation algorithm are presented in 
Simulation results are provided in \Cref{sec:simulation}.
Finally, \Cref{sec:conclusion} concludes the paper.

\textit{Notations:}
Upper-case and lower-case boldface letters denote matrices and column vectors, respectively.
The operators $(\cdot)^\T$, $(\cdot)^{*}$, and $(\cdot)^\H$ denote the transpose, complex conjugate, and Hermitian transpose, respectively.
For a vector $\mathbf{a}$, we use $| \mathbf{a} |$ to denote the Euclidean norm, and $\diag(\mathbf{a})$ to denote a diagonal matrix whose main diagonal entries are the elements of $\mathbf{a}$.
Additionally, $\mathbf{I}_n$ denotes an $n \times n$ identity matrix, and $\odot$ represents the Hadamard product.
For a complex number $z$, $\mathrm{mag}(z)$ and $\mathrm{arg}(z)$ represent the magnitude and argument of $z$, respectively.
We use $z \sim \mathcal{CN}(m,\sigma^2)$ to denote a circularly symmetric complex Gaussian random variable with mean $m$ and variance $\sigma^2$.

\section{Transceiver Architecture and Theoretical Modeling}\label{sec:system model}
% Unlike conventional wired interfacing schemes that multiplex each RF chain to antenna elements using switch networks and physical connections, the MELA system introduces an over-the-air interface between RF chains and the antenna array, as illustrated in Fig.~\ref{fig:architecture}.
% % enabling more flexible and hardware-efficient signal routing. 
% Specifically, MELA employs passive transmissive metasurfaces to relay and multiplex signals to and from sources. 
% To realize the wireless coupling mechanism, auxiliary feed antennas are deployed at the RF end to illuminate the metasurface.
% A key advantage of this approach is that phase control is delegated to a reconfigurable transmissive metasurface, eliminating the need for complex phase-shifter networks and facilitating the implementation of hybrid beamforming.
% A MELA system typically consists of a passive transmissive metasurface and a small number of active feed antennas, as schematically illustrated in Fig.~\ref{fig:architecture}.
% We consider a hybrid analog-digital transceiver architecture wherein each feed antenna connects to an RF chain, and the data streams are processed through a digital combining matrix (termed the precoding matrix in downlink transmission).

In this section, we analyze the received electric-field for the source-metasurface-receiver chain and cast it into a linear channel formulation. 
% Subsequently, we simplify and discuss the channel model under different metasurface-receiver distances.
Consider a geometric configuration in which the transmissive metasurface lies in the $yoz$ plane, with the array center positioned at the origin of the coordinate system.
The metasurface consists of $N=(2N_h+1)\times(2N_v+1)$ units, uniformly spaced by a distance $d$ in both the horizontal and vertical directions.
Assume that there are $M$ antennas at the receiver side and $K$ sources located at positions $\mathbf{p}_1, \ldots, \mathbf{p}_K$.
For simplicity, each source is modeled as an infinitesimal dipole antenna aligned along the $z$-axis.
Although infinitesimal dipoles are not physically realizable, they are commonly used to approximate short electric dipoles.
A line-of-sight path is assumed between the source and the metasurface.
Without loss of generality, we illustrate the modeling approach by focusing on the electromagnetic characteristics of the uplink channel with a point source located at $\mb{p}$, and by analyzing the power received at a transmissive unit.
Under this setting, the incident electric field at an observation point $\bar{\mathbf{t}}$ on the transmissive unit is given by
\begin{align}
    \mb{E}^i = \frac{j \eta k_c I_0 l }{4 \pi} \frac{e^{-jk_c \|\bar{\mathbf{t}} -\mb{p}\|}}{\|\bar{\mathbf{t}} - \mb{p} \|} \sin\bar{\theta} \mb{e}_{\theta}.
\end{align} 
Here, $l$ denotes the dipole length, $\eta$ the intrinsic impedance of the medium, $I_0$ the current amplitude, $j$ the imaginary unit, and $k_c=2\pi/\lambda$ the wavenumber.
We define the unit vector  $\hat{\mb{R}}=(\bar{\mathbf{t}}-\mb{p})/\|\bar{\mathbf{t}}- \mb{p}\|$, which points from the source to the observation point.
$\mb{e}_{\theta}$ is the polar unit vector in the spherical coordinate system and $\bar{\theta}$ is the angle between the radial unit vector $\hat{\mb{R}}$ and the $z$-axis.
Under the same setting, the incident magnetic field at $\bar{\mathbf{t}}$ is then given by
\begin{align}
    \mb{H}^i =\frac{1}{\eta}\hat{\mb{R}} \times \mb{E}^i
    = \frac{j k_c I_0 l }{4 \pi} \frac{e^{-jk_c \|\bar{\mathbf{t}} -\mb{p}\|}}{\|\bar{\mathbf{t}} - \mb{p} \|} \sin\bar{\theta} \mb{e}_{\phi} ,
\end{align}
where $\mb{e}_{\phi}=\hat{\mb{R}} \times \mb{e}_{\theta}$ is the azimuthal unit vector in the spherical coordinate system.

For transmissive metasurfaces, the fundamental operating mechanism is the receive-and-radiate process.
With the unit-cell transmission coefficient denoted by $\Gamma$, the transmitted magnetic field can be expressed as $\mb{H}^t = \Gamma \mb{H}^i$.
By neglecting edge effects caused by the finite aperture and applying the surface equivalence principle, each unit can be modeled as an equivalent surface current source.
Accordingly, the equivalent surface current density is given by
\begin{align}\label{eq:J}
    &\bm{\mathcal{J}}(\bar{\mathbf{t}}) = \hat{\mb{n}}\times (\mb{H}^i+\mb{H}^t)= (1+\Gamma)\mb{e}_{x} \times \mb{H}^i \notag \\
    &= \frac{(1 + \Gamma) jk_c I_0 l }{4 \pi} \frac{e^{-jk_c \|\bar{\mathbf{t}} -\mb{p}\|}}{\|\bar{\mathbf{t}}-\mb{p} \|} \sin \bar{\theta} \cos \bar{\phi}\,  \mb{e}_{z},
\end{align}
where $\hat{\mb{n}} = \mb{e}_x$ denotes the unit normal vector of the metasurface, $\bar{\phi}$ denotes the azimuth angle of $\hat{\mb{R}}$, and $\mb{e}_z$ is the unit vector along the $+z$ axis.
Unlike conventional architectures that rely on centralized phase-shifter networks, a distinguishing feature of MELA is that phase control is implemented directly at the unit-cell level.
This introduces an additional phase modulation term $e^{j\omega}$ in the equivalent surface current density, i.e., $e^{j\omega} \bm{\mathcal{J}}(\bar{\mathbf{t}})$.

Furthermore, the coordinates of any point on the transmissive unit cell can be expressed as the sum of the unit cell center position $\mathbf{t}$ and a local offset vector $\Delta \mathbf{t}$ that represents the displacement from the center (a similar decomposition applies to $\bar{\mathbf{r}}$):
\vspace{-2pt}
\begin{align*}
    \bar{\mathbf{t}}=\mathbf{t}+\Delta \mathbf{t} , \bar{\mathbf{r}}=\mathbf{r}+\Delta \mathbf{r}.
\end{align*}
\vspace{-2pt}
Throughout, we consider the practically common setting--an element-level far-field but array-level near-field regime.
Concretely, let $a_t$ and $a_r$ denote the characteristic dimensions of a transmissive unit cell and a receiving element, respectively. 
Define $R = \min \{\|\mathbf{t} - \mb{p} \|, \|\mathbf{r}- \mathbf{t}\| \}$, and assume $R \gg \max\{a_t,a_r\}$.
Under this condition, each element pair operates in the Fraunhofer region, even if the overall array-to-array link remains in the Fresnel region with respect to the array apertures.
This assumption is standard for large metasurface arrays whose unit cells are subwavelength and whose apertures are electrically large. Under these conditions, the following approximations are applied:
\begin{enumerate}
\item The field amplitudes vary negligibly across each small patch, i.e., $\|\bar{\mathbf{t}} -\mathbf{p}\|^{-1} \!\approx\! \|\mathbf{t} -\mathbf{p}\|^{-1}$ and $\|\bar{\mathbf{r}}$ $ - \mathbf{t}\|^{-1} \approx \|\mathbf{r}\!-\! \mathbf{t}\|^{-1}$, with relative errors on the order of $\mathcal O(\max\{a_t,a_r\}/R)$.
\item The direction-dependent factors are evaluated at the cell-center directions, denoted as $\sin\theta$ and $\cos\phi$, with relative errors of order $\mathcal O(\max\{a_t,a_r\}/R)$.
\item The exponential phase terms are linearized to capture the geometric phase slope across each patch, and the neglected curvature introduces a residual error of order $\mathcal{O}(k_c \max\{a_t,a_r\}^2 / R)$.
\end{enumerate}

Consequently, the electric field at the observation point $\bar{\mathbf{r}}$ due to radiation transmitted through the unit cell can be formulated as
\begin{align*}
    &\mb{E}(\bar{\mathbf{r}}) = \frac{j k_c\eta}{4\pi} \frac{e^{-j k_c \|\bar{\mathbf{r}}-\mathbf{t}\|}}{\| \bar{\mathbf{r}}-\mathbf{t}\|} \bar{\mathbf{v}} \times \Big(\bar{\mathbf{v}} \times  \int_{S_t}  e^{j \omega}  \bm{\mathcal{J}}(\bar{\mathbf{t}}) e^{j k_c \bar{\mathbf{v}} \cdot \Delta \mathbf{t} }dS \Big)  ,
    % & = \frac{j k_c\eta}{4\pi} \frac{e^{-j k_c \|\bar{\mathbf{r}}_m-\mathbf{t}_n\|}}{\| \bar{\mathbf{r}}_m-\mathbf{t}_n\|} (\mathbf{I}-\hat{\mathbf{s}}_m \hat{\mathbf{s}}_m^\T )  \int_{S_t} \!\! \bm{\mathcal{J}}(\bar{\mathbf{t}}_n) e^{j k_c \hat{\mathbf{s}}_m \cdot \bar{\mathbf{t}}_n}d\bar{\mathbf{t}}_n,
\end{align*}
where $\bar{\mathbf{v}} = (\bar{\mathbf{r}}-\mathbf{t})/ \|\bar{\mathbf{r}}-\mathbf{t}\|$ and $S_t$ represents the surface area of the  transmissive unit cell.
Considering the finite aperture of the receiving element with surface area $S_r$, the average electric field over the receiving element can be expressed as
\begin{align} \label{eq:E_rm}
\bar{\mathbf{E}}_{r} = \frac{1}{S_r} \int_{S_r} \mathbf{E}(\bar{\mathbf{r}}) dS \approx  \frac{\mathbf{E}(\mathbf{r}) }{S_r} \int_{S_r}  e^{-j k_c \hat{\mathbf{v}} \cdot \Delta \mathbf{r} }dS,
\end{align}
where $\hat{\mathbf{v}} = (\mathbf{r}-\mathbf{t})/ \|\mathbf{r}-\mathbf{t}\|$ is the propagation direction from the center of unit cell to the center of receiving element, and
\begin{align} \label{eq:E_r}
    \!\!\!\!\mathbf{E}(\mathbf{r}) \!=\! \frac{j k_c\eta}{4\pi} \frac{e^{-j k_c \|\mathbf{r}-\mathbf{t}\|}}{\|\mathbf{r}-\mathbf{t}\|} \hat{\mathbf{v}} \!\times\!\! \Big(\hat{\mathbf{v}}\! \times \!\! \int_{S_t} \!\! e^{j \omega}  \bm{\mathcal{J}}(\bar{\mathbf{t}}) e^{j k_c \hat{\mathbf{v}}\cdot \Delta \mathbf{t} }dS \Big) .
\end{align}
Substituting \eqref{eq:J} and \eqref{eq:E_r} into \eqref{eq:E_rm} yields the final expression, denoted as \eqref{eq:E-int}.
Furthermore, if the apertures have regular and analytically tractable geometries, such as squares, the expression can be further simplified, leading to the following theorem.

\begin{figure*}[t]
    \begin{equation} \label{eq:E-int}
        \begin{split}
            \bar{\mathbf{E}}_{r}  =& \frac{(1+\Gamma)k_c^2 \eta I_0 l}{(4 \pi)^2 S_r } \frac{e^{-jk_c \| \mathbf{r}-\mathbf{t} \|} e^{-j k_c\|\mb{t}-\mathbf{p}\| }}{\|\mathbf{r}-\mathbf{t}\| \|\mb{t}-\mathbf{p} \|} \int_{S_t}  e^{j k_c (\frac{\mathbf{r}-\mathbf{t} }{\| \mathbf{r}-\mathbf{t} \|}-\frac{\mathbf{t}-\mb{p} }{\|\mathbf{t}-\mb{p} \|})^\T  \Delta \mathbf{t} } dS \int_{S_r} e^{-j k_c\frac{(\mathbf{r}-\mathbf{t})^\T  }{\| \mathbf{r}-\mathbf{t} \|} \Delta \mathbf{r} } dS  \sin \theta \cos \phi e^{j \omega} (\mathbf{I}-\hat{\mathbf{v}}\hat{\mathbf{v}}^\T) \mathbf{e}_z.
        \end{split}
    \end{equation}
    \vspace{-15pt}
\end{figure*}

\begin{Theorem} \label{theo:main}
Suppose that both the receiving antenna and the transmissive unit cell have square apertures with side lengths $d_r$ and $d_t$. 
Let $\mathbf{e}_{\text{co}}$ be the receiver's polarization unit vector.
Then, the average electric field at the $m$-th receiving antenna, due to a source located at position $\mathbf{p}_k$ after propagation through the $n$-th transmissive unit cell, is given by
\begin{align} 
    \label{eq:theo1}
    \bar{\mb{E}}_r = \mathcal{A}_{mn} \mathcal{B}_{mnk}\frac{e^{-j k_c \|\mathbf{r}_m-\mathbf{t}_n\|} e^{-j k_c \|\mathbf{t}_n-\mb{p}_k\|} }{\|\mathbf{r}_m-\mathbf{t}_n\| \|\mathbf{t}_n-\mb{p}_k\|}  e^{j \omega_n} I_0 l ,
\end{align}
where the scalars $\mathcal{A}_{mn}$ and $\mathcal{B}_{mnk}$ are defined as
\begin{align*}
    &\mathcal{A}_{mn} = \frac{k_c \eta}{4 \pi}  \sinc\big( \frac{k_c v_y d_r}{2} \big) \sinc\big( \frac{k_c v_z d_r}{2} \big)\mathbf{e}_{\text{co}}^\H(\mathbf{I}-\hat{\mb{v}}_{mn}\hat{\mb{v}}_{mn}^\T) \mathbf{e}_z, \\
    &\mathcal{B}_{mnk} \!= \!\frac{(1\!+\!\Gamma)k_c d_t^2 }{4 \pi}  \sinc\!\big( \frac{k_c u_y d_t}{2}  \big) \!\sinc\!\big( \frac{k_c u_z d_t}{2}  \big) \sin \theta_{\!nk}\! \cos \phi_{\!nk}.
\end{align*}
% 坐标系定义混乱了，这里的角度和后面球坐标中的角度定义混了。
Here, $\theta_{nk}$ and $\phi_{nk}$ denote the azimuth and elevation angles of the vector $\mathbf{t}_n - \mathbf{p}_k$.
The direction vectors $\hat{\mb{u}}_{nk}$ and $\hat{\mb{v}}_{mn}$ are defined as
\begin{align*}
    &\hat{\mb{v}}_{mn} = (\mathbf{r}_m-\mathbf{t}_n)/\|\mathbf{r}_m-\mathbf{t}_n \|= [v_x,v_y,v_z]^\T, \\
    &\hat{\mb{u}}_{mnk} = (\mathbf{t}_{n}-\mb{p}_k )/\|\mathbf{t}_{n}-\mb{p}_k \|-\hat{\mb{v}}_{mn} = [u_x,u_y,u_z]^\T. 
\end{align*}
% The detailed derivation is provided in \Cref{append} of the appendix.
\end{Theorem}

\begin{proof}
The sinc factors are obtained by integrating the exponential phase term over the square aperture of each element.
Separating the surface integral along the $y$- and $z$-axes yields two one-dimensional terms $\frac{2\sin(\tfrac{k_c u_{y,z} d_t}{2})}{k_c u_{y,z}}$, whose product forms the final sinc functions.
\end{proof}

Theorem~\ref{theo:main} derives a closed-form expression for the amplitude of the electric field at the receiving element.
The sinc functions in $\mathcal{A}_{mn}$ and $\mathcal{B}_{mnk}$ represent the directional responses of the receiving apertures and the transmissive unit cells, respectively. 
When $d_r \ll \lambda$ or $d_t \ll \lambda$, the corresponding sinc terms approach unity, implying that the receiving aperture or the transmissive unit cell behaves approximately isotropically.
To build the complete propagation model, we project the vector field onto each receiver's co-polarization and apply the field-to-port scaling $\sqrt{A_{\text{eff,m}}/2\eta}$, where $A_{\text{eff,m}}$ represents the effective aperture of the $m$-th receiver.
We then invoke linear superposition to sum the contributions of all unit cells and thereby convert the EM-field description into a linear  baseband channel model, which yields the following corollary.

\begin{Corollary}\label{theo:MatrixModel}
Assume there are $K$ sources with infinitesimal dipole antennas, and consider a digital decoding architecture with $M$ RF chains connected to $M$ receiver antennas.
Let $\mathbf{H}=[\mathbf{h}_1,\ldots,\mathbf{h}_M]^\T \in \mathbb{C}^{M \times N}$ denote the channel matrix characterizing the propagation from the transmissive metasurface to the receiving antennas, with each element given by $ [\mathbf{H}]_{m,n} \!=\!\sqrt{\frac{A_{\text{eff,m}}}{2\eta}}\,  \mathcal{A}_{mn} \frac{e^{-jk_c \| \mathbf{r}_m-\mathbf{t}_n \|}}{\| \mathbf{r}_m-\mathbf{t}_n\|} $.
% \begin{align} \label{eq:E_Hmn}
%     [\mathbf{H}]_{m,n} =\sqrt{\frac{A_{\text{eff,m}}}{2\eta}}\,  \mathcal{A}_{mn} \frac{e^{-jk_c \| \mathbf{r}_m-\mathbf{t}_n \|}}{\| \mathbf{r}_m-\mathbf{t}_n\|}  .
% \end{align}
Let $\mathbf{G}_m \!\in\! \mathbb{C}^{N \times K}$ denote the effective source-metasurface coupling as perceived by the $m$-th receive antenna, with each element given by $ [\mathbf{G}_m]_{n,k} =
    \mathcal{B}_{mnk} \frac{e^{-j k_c \|\mathbf{t}_{n}-\mathbf{p}_k\|}}
         {\|\mathbf{t}_{n}-\mathbf{p}_k\|}$.
% \begin{equation}\label{eq:k_column}
%     [\mathbf{G}_m]_{n,k} =
%     \frac{\mathcal{B}_{mnk}\, e^{-j k_c \|\mathbf{t}_{n}-\mathbf{p}_k\|}}
%          {\|\mathbf{t}_{n}-\mathbf{p}_k\|}.
% \end{equation}
Accordingly, the received baseband signal after combining can be expressed as
\begin{align} \label{eq:channel model}
    \mathbf{y}
    = \textstyle\sum_{m=1}^{M} \mathbf{w}_m
      \big( \mathbf{h}_m^\T \bPhi \mathbf{G}_m \mathbf{s} + n_m \big),
\end{align}
where $\mathbf{s}=[I_{0,1}l_1 s_1,\ldots,I_{0,K}l_K s_K]^\T \in \mathbb{C}^K$
denotes the equivalent transmitted signal vector, with $s_k$ representing the $k$-th information symbol.
The matrix $\mathbf{W}_{\text{BB}}=[\mathbf{w}_1^\H,\ldots,\mathbf{w}_M^\H]\in\mathbb{C}^{M \times N_s}$ is the digital combining matrix mapping $M$ RF chains to $N_s$ data streams.
The noise $n_m \sim \mathcal{CN}(0,\sigma^2)$ denotes the additive white Gaussian noise at the $m$-th antenna, with $\sigma^2$ being the noise power.
Finally, $\bPhi=\mathrm{diag}(\bm{\omega})\in\mathbb{C}^{N\times N}$ represents the metasurface phase configuration,
where $\bm{\omega}=[e^{j\omega_1},\ldots,e^{j\omega_N}]^\T$ contains the unit-modulus phase shifts of the transmissive elements.
\end{Corollary}

Corollary~\ref{theo:MatrixModel} provides a physically interpretable and mathematically tractable formulation that explicitly captures the effects of spatial geometry, phase configuration, and wave propagation. 
The channel matrices $\mathbf{H}$ and ${\mathbf{G}_m}$ represent the complex-valued propagation responses along the paths from the transmissive metasurface to the receiver array and from the sources to the metasurface, respectively.
Notably, these matrices exhibit well-defined structures under certain geometric or operational conditions.

\begin{figure}[t]
    \centering
    \includegraphics[width=0.9\linewidth]{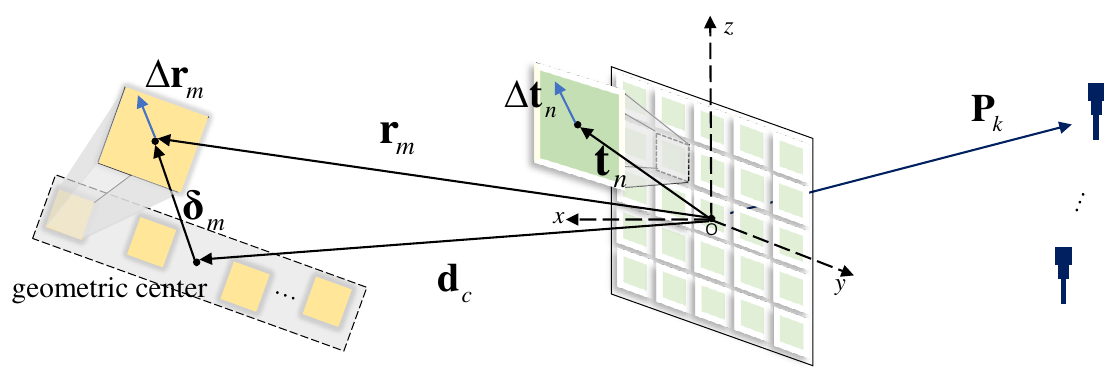}
    \caption{Diagram of the transceiver architecture with receiving antennas confined to a localized region.}\label{fig:array element}
    \vspace{-5pt}
\end{figure}

One of the most common practical cases is when the receiving antennas are co-located or closely spaced within a compact region, for example arranged as a uniform linear array.
Consider the case of co-located receiving apertures and define $\| \mathbf{d}_{mn} \| \triangleq \|\mathbf{r}_m - \mathbf{t}_n\|$.
Since the spatial variation across the receiving elements is limited, the exponential phase term can be approximated by a second-order Taylor expansion around the array center.
Specifically, the position of the $m$-th antenna element can be expressed as
\begin{equation} \label{eq:def_rm}
    \vspace{-2pt}
    \mathbf{r}_m = \mathbf{d}_c + \bm{\delta}_m ,
    \vspace{-2pt}
\end{equation}
where $\mathbf{d}_c$ denotes the geometric center of the receiving array, and $\bm{\delta}_m$ represents the relative displacement vector from $\mathbf{d}_c$ to the $m$-th antenna element, as illustrated in Fig.~\ref{fig:array element}.
Substituting \eqref{eq:def_rm} into $\| \mathbf{d}_{mn} \|$ and using the second-order Taylor expansion, the expression of $\| \mathbf{d}_{mn} \|$ can be approximated as
\begin{align} \label{eq:dmn}
& \! \| \mb{d}_{mn} \| \!
\approx\! \| \mathbf{d}_c \| \!+ \! \frac{\mathbf{d}_c^\T (\bm{\delta}_m \!-\! \mathbf{t}_{n} )}{\| \mathbf{d}_c \|} \! + \!\frac{(\bm{\delta}_m \!-\! \mathbf{t}_{n})^\T \!\mathcal{H}(\bm{\delta}_m \!-\! \mathbf{t}_{n})}{2} \notag \\
& \!=\! \| \mathbf{d}_c \| \!+\! \frac{\mathbf{d}_c^\T (\bm{\delta}_m \!-\! \mathbf{t}_{n})}{\| \mathbf{d}_c \|} \!+\! \frac{\bm{\delta}_m^\T \mathcal{H} \bm{\delta}_m \!+\! \mathbf{t}_{n}^\T \mathcal{H}\mathbf{t}_{n}}{2} \!-\! \mathbf{t}_n^\T \mathcal{H}\bm{\delta}_m ,
\end{align}
where $\mathbf{\mathcal{H}}$ denotes the Hessian matrix given by $\mathbf{\mathbf{\mathcal{H}}} = (\| \mathbf{d}_c \|^2 \mathbf{I}- \mathbf{d}_c  \mathbf{d}_c ^\T)/\|  \mathbf{d}_c  \|^3$.
Similarly, the following approximations hold
\begin{align*}
    &\|  \mathbf{d}_c+\bm{\delta}_m \|=  \|  \mathbf{d}_c \| + \mathbf{d}_c^\T \bm{\delta}_m/\|  \mathbf{d}_c \| + \bm{\delta}_m^\T \mathbf{\mathcal{H}}\bm{\delta}_m/2 +o(\| \bm{\delta}_m\|^2), \\
    &\| \mathbf{d}_c-\mathbf{t}_{n} \|=  \| \mathbf{d}_c \| -  \mathbf{d}_c^\T \mathbf{t}_n/\| \mathbf{d}_c \| + \mathbf{t}_n^\T \mathbf{\mathcal{H}}\mathbf{t}_n/2 +o(\|  \mathbf{t}_n\|^2).
\end{align*}
Neglecting higher-order terms, the expression in \eqref{eq:dmn} can be rewritten as
\begin{align} \label{eq:after}
\| \mb{d}_{mn}  \| \approx & \| \mathbf{d}_c+\bm{\delta}_m \|+\| \mathbf{d}_c-\mathbf{t}_{n} \|-\|\mathbf{d}_c\| - \mathbf{t}_n^\T \mathbf{\mathcal{H}}\bm{\delta}_m.
\end{align}

Comparing \eqref{eq:dmn} and \eqref{eq:after}, we observe that, unlike the original non-negative quadratic form in \eqref{eq:dmn}, the bilinear cross term in \eqref{eq:after} generally has smaller magnitude and may even vanish, thus introducing only a minor perturbation to the overall approximation.
When the receiving antennas are co-located within a confined region such that the cross term $\mathbf{t}_n^\T \mathcal{H} \bm{\delta}_m$ is sufficiently small, 
% --for example, when the displacement vector $\bm{\delta}_m$ is small in magnitude--
the distance term $\| \mb{d}_{mn}  \|$ can be approximated as 
\begin{align} \label{eq:decoupled-distance}
     \| \mb{d}_{mn}  \| \approx & \| \mathbf{d}_c+\bm{\delta}_m \|+\| \mathbf{d}_c-\mathbf{t}_{n} \|-\|\mathbf{d}_c\| .
\end{align}
% This expression decouples the influence of the $n$-th transmissive unit cell position and the $m$-th receiver position. 
We refer to this quantity as the approximately decoupled distance, as it separates the contributions of the receiver and transmissive unit cell geometries relative to their respective array centers.

To further elucidate the validity of the decoupling approximation, we consider a representative case where the receiving antenna array is configured as a uniform linear array (ULA) aligned along the $y$-axis and placed in the $z=0$ plane.
In this configuration, the maximum phase discrepancy between the exact distance expression in \eqref{eq:dmn} and its decoupled approximation in \eqref{eq:decoupled-distance} can be bounded as
\begin{equation} \label{eq:mu1}
   \left\lvert \frac{2\pi}{\lambda} \left( \frac{\mathbf{t}_{n}^\T \bm{\delta}_m}{\|\mathbf{d}_c\|}-\frac{\mathbf{t}_{n}^\T \mathbf{d}_c \mathbf{d}_c^\T \bm{\delta}_m}{\|\mathbf{d}_c\|^3} \right) \right\rvert \le \frac{2\pi D_1 D_h}{\lambda \| \mathbf{d}_c \|} \le \epsilon,
\end{equation}
where $D_1$ denotes the maximum aperture diameter of the receiving array, $D_h=(2N_h + 1)d$ represents the physical apertures of the metasurface along the horizontal axes, and $\epsilon$ is the prescribed phase error tolerance. This phase discrepancy attains its maximum when $\mathbf{t}_{n} = [0, D_h/2, 0]^\T$ and $\bm{\delta}_m = [0, D_1/2, 0]^\T$. The corresponding minimum distance required to ensure the approximation remains within the specified error bound is given by
\begin{equation}\label{E:Bound_1}
R_{\epsilon}^d = 2 \pi D_1 D_h/\lambda \epsilon .
\end{equation}
Thus, when $\| \mathbf{d}_{c} \| > R^d_{\epsilon}$, the approximation error becomes negligible. For instance, by choosing $\epsilon = \pi / 8$, we obtain $R^d_{\epsilon} = 4 D_1 D_h/\lambda $.
% \[  %\label{eq:R1}
% R^d_{\epsilon} = 4 D_1 D_h/\lambda .
% \]

% With the decoupled distance approximation in \eqref{eq:decoupled-distance}, each entry of the channel matrix $\mathbf{H}$ can be expressed as
% \begin{equation*}
%     \mathbf{H}_{mn}\!=\!\sqrt{\!\frac{A_{\text{eff,m}}}{2\eta}} \frac{\mathcal{A}_{mn}}{\| \mb{d}_{mn} \|} e^{j k_c \|\mathbf{d}_c\|} e^{-j k_c \|\mathbf{d}_c+\bm{\delta}_m \|} e^{-j k_c -\|\mathbf{d}_c-\mathbf{t}_n \|}.
% \end{equation*} 
Furthermore, under the distance condition $\|\mathbf{d}_c\| > R_{\epsilon}^d$, the approximation 
$\mathcal{B}_{mnk}\!\approx\!\mathcal{B}_{nk}$ introduces only negligible modeling error while significantly simplifying the channel representation (see Appendix \ref{appd:A} for the detailed derivation).  
This leads to the following corollary.
\begin{Corollary}\label{cor:factorized_H}
Under the decoupled distance approximation, the source-metasurface channel matrix $\mathbf{G}_m$ becomes independent of the receive index $m$ and is hence denoted by $\mathbf{G}$.  
The metasurface-receiver channel matrix $\mathbf{H}$ admits the factorized form $\mathbf{H} =\diag(\mb{h}_r) \mathbf{C} \diag(\mb{h}_t)$, 
% \begin{equation} \label{eq:H_decoupled}
%     \mathbf{H} =\diag(\mb{h}_r) \mathbf{C} \diag(\mb{h}_t) ,
% \end{equation}
where $\diag(\mb{h}_r)$ and $\diag(\mb{h}_t)$ characterize the spatial phase responses at the receiver and transmissive metasurface sides, respectively, and the matrix $\mathbf{C}$ represents amplitude attenuation. The components are defined as
\begin{align*}
    &\mb{h}_r= [e^{-jk_c \| \mathbf{d}_c+\bm{\delta}_1 \|},\ldots, e^{-jk_c \| \mathbf{d}_c+\bm{\delta}_M \|}]^\T , \\
    &\mb{h}_t= [e^{-jk_c \| \mathbf{d}_c-\mathbf{t}_{1} \|},\ldots, e^{-jk_c \| \mathbf{d}_c-\mathbf{t}_{N} \|}]^\T , \\
    &[\mathbf{C}]_{mn}= \sqrt{A_{\text{eff,m}}/2\eta} \,  \mathcal{A}_{mn} e^{j k_c \| \mathbf{d}_c \| }/ \| \mb{d}_{mn} \|.
\end{align*}
Accordingly, the received baseband signal can be written in the compact linear matrix form
\begin{align} \label{eq:H_decoupled}
    \mathbf{y} = \mathbf{W}_{\text{BB}}^\H \mathbf{H} \bPhi \mb{G} \mathbf{s} + \mb{n},
\end{align}
where the noise vector $\mb{n}=\mathbf{W}_{\text{BB}}^\H [n_1,\ldots,n_M]^\T \in \mathbb{C}^M$.
\end{Corollary}

\begin{figure}[t]
    \centering
    \includegraphics[width=0.95\linewidth]{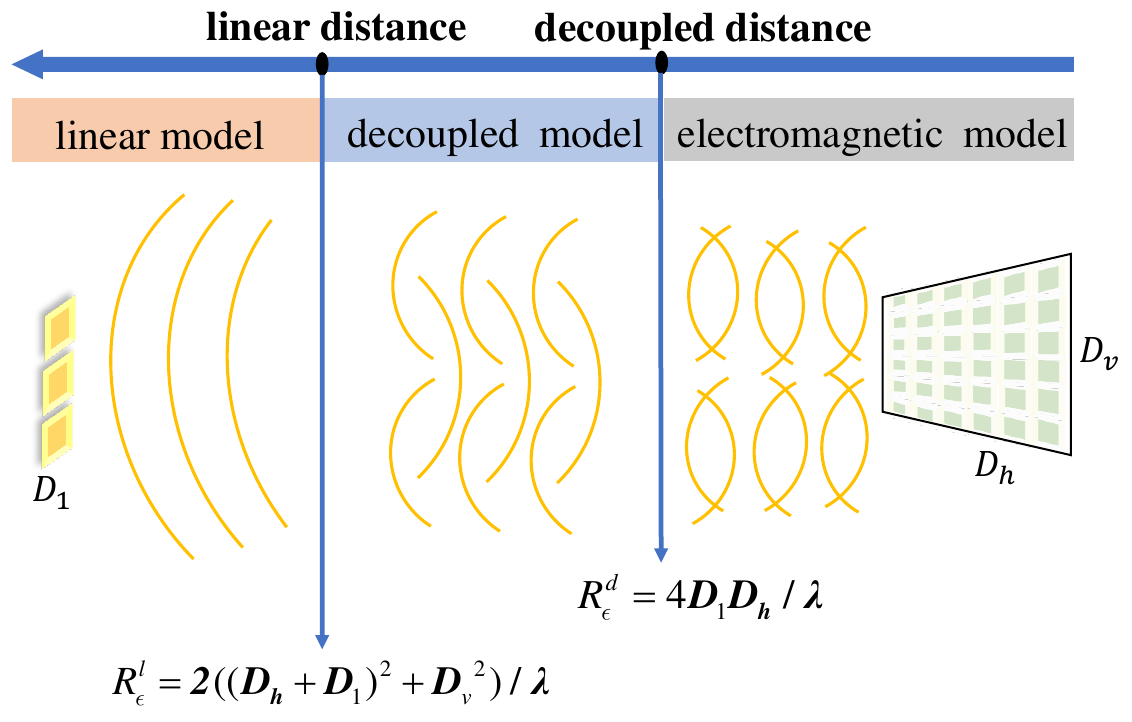}
    \caption{Illustration of different modeling regimes based on the metasurface-receiver distance.}\label{distance}
    \vspace{-5pt}
\end{figure}

% \subsection{Far-field}
If, in addition, the non-negative quadratic form in \eqref{eq:dmn} can be neglected, such as when $\| \bm{\delta}_m-\mathbf{t}_{n} \|$ is sufficiently small relative to the distance $\| \mathbf{d}_c \|$, the distance approximation further simplifies. 
In this case, the second-order geometric effects become negligible, and the path length $\|\mathbf{d}_{mn}\|$ is dominated by the first-order terms, yielding
\begin{align} \label{eq:far-distance}
    \| \mb{d}_{mn}  \| \approx  \| \mathbf{d}_c \| + \mathbf{d}_c^\T (\bm{\delta}_m-\mathbf{t}_{n})/\| \mathbf{d}_c \|.
\end{align}
To quantify the validity of this linear approximation, we evaluate the phase discrepancy between the exact expression in \eqref{eq:dmn} and its linearized counterpart in \eqref{eq:far-distance}. 
Similar to the previous discussion, and assuming the receiving antenna array is configured as a uniform linear array, the resulting phase error is bounded by
\begin{equation}
\!\!\! \left \lvert \frac{2\pi}{\lambda} \frac{(\bm{\delta}_m\!-\!\mathbf{t}_{n})^\T \mathcal{H}(\bm{\delta}_m\!-\!\mathbf{t}_{n})}{2} \right \rvert \!\le\! \frac{ \pi ( ( D_h\!+\!D_1 )^2 \!+\! D_v^2 ) }{4 \lambda \| \mathbf{d}_c \|} \!\le\! \epsilon ,
\end{equation}
where $\epsilon$ denotes the prescribed phase error tolerance. This leads to the following minimum distance requirement to ensure the validity of the linear approximation
\begin{equation}\label{E:Bound_2}
R_{\epsilon}^l = \pi ( (D_h+D_1)^2+D_v^2 )/(4 \lambda \epsilon) ,
\end{equation}
% Under this configuration, the physical apertures of the array along the horizontal and vertical axes are given by $D_h=(2N_h + 1)d$ and $D_v=(2N_v + 1)d$, respectively.
where $D_v=(2N_v + 1)d$ represents the physical apertures of the metasurface along the vertical axes.
When $\| \mathbf{d}_c \| \ge R_{\epsilon}^l$, the approximation error becomes negligible.
Comparing the bounds in \eqref{E:Bound_1} and \eqref{E:Bound_2}, we observe that the minimum distance required for the validity of the decoupled approximation $R_{\epsilon}^d$, is generally more stringent than that for the linear approximation, $R_{\epsilon}^l$, as shown in Fig.~\ref{distance}. 
This indicates that the decoupled model remains accurate across a wider range of practical configurations, while the linear model demands a stricter condition to limit the phase error within $\epsilon$. For example, by choosing $\epsilon = \pi / 8$, we have $R_{\epsilon}^l = 2((D_h+D_1)^2+D_v^2)/\lambda$, which clearly exceeds $R_{\epsilon}^d$.
% \begin{align} \label{eq:def Rayleigh distance}
%  R_{\epsilon}^l = \frac{2((D_h+D_1)^2+D_v^2)}{\lambda} ,
% \end{align}
% which exceeds $R_{\epsilon}^d$.

With the distance approximation in \eqref{eq:far-distance}, the channel matrix $\mathbf{H}$ takes the following simplified form:
\begin{align} \label{eq:H_far}
    \mathbf{H}= \sqrt{\frac{A_{\text{eff,m}}}{2\eta}}  \mathcal{A}_c \frac{e^{-j k_c \|\mathbf{d}_c\|}}{ \| \mb{d}_c \|} \mb{h}_M \mb{h}_N^\T .
\end{align}
Here, $\mb{h}_M$ and $\mb{h}_N$ denote the steering vectors, defined as
\begin{align*}
    &\mb{h}_M=[e^{-jk_c\mathbf{d}_c^\T \bm{\delta}_1 / \|\mathbf{d}_c\| },\ldots, e^{-jk_c\mathbf{d}_c^\T \bm{\delta}_M / \|\mathbf{d}_c\| }]^\T, \\
    &\mb{h}_N = [e^{jk_c \mathbf{d}_c^\T \mathbf{t}_1 /\|\mathbf{d}_c\|},\ldots,e^{jk_c \mathbf{d}_c^\T \mathbf{t}_N /\|\mathbf{d}_c\|} ]^\T.
\end{align*}
The term $\mathcal{A}_c$ denotes the attenuation coefficient, expressed as
\begin{align*}
    \mathcal{A}_c =  \frac{k_c \eta  }{4 \pi} 
     \sinc \big( \frac{k_c d_r d_y}{2}  \big) \sinc \big( \frac{k_c d_r d_z}{2}  \big) \mathbf{e}_{\text{co}}(\mathbf{I}-\mb{d}_c\mb{d}_c^\T) \mathbf{e}_z,
\end{align*}
where $d_y$ and $d_z$ denote the $y$- and $z$-components of $ \mb{d}_c/\|\mb{d}_c \|$, respectively.
    Consider a practical MELA system operating at a carrier frequency of $60$~GHz, corresponding to a wavelength of $0.5$~cm. 
    The metasurface consists of $21 \times 21$ elements, and the receiver array contains $15$ elements, with both having a unit spacing of half-wavelength. 
    Consequently, the physical aperture of the receiver array is $3.75$~cm, and the horizontal and vertical apertures of the metasurface are both $5.25$~cm.
    Under this configuration, by choosing $\epsilon=\pi/8$, the values of $R_{\epsilon}^d$ and $R_{\epsilon}^l$ are $1.58$~m and $4.34$~m, respectively.
% \end{Remark}

% \begin{Remark} \label{rem:symmetric}
%     In the downlink scenario, the same surface currents on the feed antenna at $\mathbf{r}_m$ radiate back to the dipole at $\mathbf{p}$ through the transmissive patch antenna $\mathbf{t}_n$.
%     In both directions we end up with the same path-loss factor the same phase shifts imposed by the surface currents.
%     By the Lorentz reciprocity theorem, for time-invariant, linear media, the Green function from point A to point B equals that from B to A.
%     Concretely, it is easy to check that the uplink and downlink channels share an identical (or symmetric) algebraic form.
% \end{Remark}

% \newpage
\section{Channel estimation}
\label{sec:channel es}

In this section, we focus on uplink channel estimation of the MELA system, which is a necessary step for subsequent analog/digital precoding in conventional communication pipelines.
We consider a hybrid-field scenario where some sources reside in the near-field region, while others are in the far-field.
At moderately close range, the phase varies significantly across each unit cell, whereas the wavefront amplitude remains approximately uniform \cite{calvez2018massive}.
Accordingly, we adopt a phase-only array-manifold model, which suffices for reliable estimation in both near- and far-field regimes.
% Consequently, a phase-only manifold is adequate for near- and far-field estimation.
With this model, the $k$-column of the matrix $\mathbf{G}$ can be expressed as
\begin{align*}
    \mb{G}_{:,k} = \left[ \frac{\mathcal{B}_{k}}{\|\mb{p}_k\|}e^{j k_c \|\mb{p}_k-\mathbf{t}_1\|},\ldots, \frac{\mathcal{B}_{k}}{\|\mb{p}_k\|}e^{j k_c \|\mb{p}_k-\mathbf{t}_N\|} \right]^\T ,
\end{align*}
where the subscript $n$ in coefficient $\mathcal{B}_{nk}$ is omitted.
%  since the amplitude variation across array elements is negligible.
% , where $\theta_k$ and $\phi_k$ denote the source's azimuth and elevation angles relative to the center of the metasurface, respectively.
% Here, the elevation angle is defined as the angle between the source and the $x$o$y$ plane, while the azimuth angle refers to the angle between the source and the $x$o$z$ plane.
The coordinates of the $(n_y, n_z)$-th element on the metasurface are given by $[0,\ n_y d,\ n_z d ]^\T$, where $d$ is the inter-element spacing.
We represent the position of the $k$-th source as $\mathbf{p}_k = r_k [\sqrt{\cos^2 \phi_k - \sin^2 \theta_k},\ \sin \theta_k,\ \sin \phi_k ]^\T$.
Substituting coordinates, the matrix $\mathbf{G}$ can be expressed in terms of spherical wave steering vectors as
\begin{align} \label{eq:G}
    \mathbf{G} = [\frac{\mathcal{B}_1}{r_1} \bm{\alpha}(\theta_1,\phi_1,r_1),\ldots,\frac{\mathcal{B}_K}{r_K} \bm{\alpha}(\theta_K,\phi_K,r_K) ] , 
\end{align}
where $\bm{\alpha}(\theta, \phi, r) = [e^{j k_c d_1}, \ldots, e^{j k_c d_N}]^\T$.
In the near-field, the source-to-$n$th metasurface element distance $d_n$ admits the following second-order expansion
\begin{align} \label{eq:dn_near}
        d_n &=\sqrt{r^2-2r\sin\theta n_y d-2r\sin\phi n_z d +n_y^2 d^2+n_z^2 d^2}  \notag\\
        & \approx r-d(\sin\theta n_y+\sin\phi n_z) - \frac{d^2 }{r} \sin\theta \sin\phi n_y n_z \notag\\
    &\quad + \frac{d^2}{2r}(\cos^2\theta n_y^2 + \cos^2 \phi n_z^2) .
\end{align}

\begin{figure*}[t]
    \centering
    % \vspace{15pt}
    \begin{minipage}{\linewidth}
        % \hrule
    \begin{align} \label{eq:long-st-vector}
        \bm{\alpha}(\gamma^a,\gamma^e,& \beta^a,\beta^e,\alpha)\! =[ e^{j(-N_h \gamma^a-N_v\gamma^e+(-N_h)^2\beta^a+(-N_v)^2 \beta^e +(-N_h)(-N_v)\alpha)},\!\ldots\!,e^{j(-N_h \gamma^a+N_v\gamma^e+(-N_h)^2\beta^a+N_v^2 \beta^e +(-N_h)N_v\alpha)}, \notag \\
        &\!\! e^{j((-N_h+1) \gamma^a-N_v\gamma^e+(-N_h+1)^2\beta^a+(-N_v)^2 \beta^e +(-N_h+1)(-N_v)\alpha)}, \!\ldots\!,e^{j(N_h \gamma^a+N_v\gamma^e+N_h^2\beta^a+N_v^2 \beta^e +N_h N_v\alpha)}]^\T  e^{j k_c r} .
    \end{align}
    \end{minipage}
\vspace{-4pt}
\end{figure*}

% When the communication distance falls within the Fresnel region, a second-order Taylor expansion provides a sufficiently accurate approximation for the distance term.
% In this regime, the distance $d_n$ can be approximated as
% \begin{equation} \label{eq:dn_near}
%     \begin{split}
%     d_{n} \approx &r-d(\sin\theta n_y+\sin\phi n_z) - \frac{d^2 \sin\theta \sin\phi n_y n_z}{r}\\
%     &+ \frac{d^2}{2r}(\cos^2\theta n_y^2 + \cos^2 \phi n_z^2) .
%     \end{split}
% \end{equation}
% \begin{equation} \label{eq:dn_near}
%     \begin{split}
%     d_{n} \approx &r-d(\sin\theta n_y+\sin\phi n_z) - d^2 \sin\theta \sin\phi n_y n_z/r\\
%     &+ d^2(\cos^2\theta n_y^2 + \cos^2 \phi n_z^2)/(2r) .
%     \end{split}
% \end{equation}

In the far-field case, where $r$ is typically much larger than the aperture of metasurface, the second-order terms become negligible, and the distance expression simplifies to
\begin{align} \label{eq:r-far}
   d_n \approx  r-d(\sin\theta n_y+\sin\phi n_z).
\end{align}
We define the following parameters:
\begin{align*}
    &\gamma^a = -k_c d \sin \theta,\gamma^e =- k_c d \sin \phi,   \\
    & \beta^e = \frac{k_c d^2 \cos^2 \phi}{2 r}, \beta^a = \frac{k_c d^2 \cos^2 \theta}{2 r},\alpha = -\frac{k_c d^2 \sin \theta \sin\phi}{r},
\end{align*}
and substitute them into the distance approximation in \eqref{eq:dn_near}.
This yields the near-field steering vector $\bm{\alpha}(\gamma^a,\gamma^e, \beta^a,\beta^e,\alpha)$, as expressed in \eqref{eq:long-st-vector}.
In the far-field regime, the second-order terms vanish, i.e., $ \beta^a = \beta^e = \alpha = o(1)$, and the steering vector simplifies to the planar-wave form
\begin{align*}
    \bm{\alpha}(\gamma^a,\gamma^e)= [& e^{j(-N_h \gamma^a-N_v\gamma^e)},\ldots,e^{j(-N_h \gamma^a+N_v\gamma^e)},\\
    & e^{j((-N_h+1) \gamma^a-N_v\gamma^e)}, \ldots,e^{j(N_h \gamma^a+N_v\gamma^e)}]^\T  e^{j k_c r}.
\end{align*}

By substituting \eqref{eq:G} into \eqref{eq:channel model}, the received signal can be written as:
\begin{align}
\mathbf{y}= \mathbf{W}_{\text{BB}}^\H \mathbf{H}\bm{\Phi} \textstyle \sum_{k=1}^K \frac{\mathcal{B}_k}{r_k} \bm{\alpha}(\theta_k,\phi_k,r_k) \mathbf{s}(k) + \mathbf{n},
\end{align}
where the channel matrix $\mathbf{H}$ is assumed to be known.
In the subsequent channel estimation procedure, we set the digital combiner to the identity, i.e., $ \mathbf{W}_{\text{BB}}= \mathbf{I}_M $, so that each RF chain is passed through without any inter-chain mixing during training.
We first apply a dictionary-based beamspace filtering technique (see \Cref{subsec:beam_filter}) to obtain a coarse estimation in the angular domain.
Based on the coarse angular sets, we then perform refined estimation of angle and distance as described in \Cref{subsec:refined_es}.

\vspace{-5pt}
\subsection{Coarse Estimation}\label{subsec:beam_filter}
When the channel varies slowly or high estimation accuracy is not strictly required, the rapid phase-switching capability of metasurface can be leveraged to perform full angular space scanning. 
The core idea is to dynamically configure the phase coefficients of the metasurface across time slots, thereby constructing a measurement function that maps angular directions to received signal energy. 
We refer to this approach as dictionary-driven beamspace filtering, which combines two core mechanisms: 1) a metasurface-generated beamspace dictionary to sparsify the channel response, and 2) a filtering-inspired reconstruction to extract angular information. 

For analytical tractability, we denote the number of sampling points as $T_1 = P \times Q$, where $P$ and $Q$ represent the quantization levels for the azimuth and elevation angles, respectively.
During the scanning phase, it is assumed that the symbols transmitted by sources remain constant.
At sub-slot $t = p \times q$, the phase shift of the $n$-th unit cell with row-column indices $(n_h,n_v)$ is configured to simultaneously compensate for the phase delays toward both the receiving elements and the discretized angular direction $(\theta_p, \phi_q)$:
\begin{align} \label{eq:phase_selection}
    \omega_n&=k_c d (\sin\theta_p n_h+\sin\phi_q n_v)+ \arg(h_n ) \notag \\
    &= \arg(\bm{\alpha}^*(\theta_p,\phi_q,\infty)) + \arg(h_n ),
\end{align}
where $h_n = \mathrm{sum}( \mathbf{H}_{[:,n]}) $ denotes the sum of the entries in the $n$-th column of the matrix $\mathbf H$.
Subsequently, the received signals at each sub-slot $t$ are summed to construct the following measurement function:
\begin{align} \label{eq:measurement_func}
    f_t & \triangleq  \mathrm{sum}(\mb{y}_t) =[h_1, h_2, \ldots, h_N] \bPhi_t \mathbf{G} \mathbf{s} + \mathrm{sum}(\mb{n}_t).
\end{align}
% where $\mathbf{h} = [h_1, \ldots, h_N]^\T \in \mathbb{C}^N$ denotes the aggregated metasurface-to-receiver channel.
The phase configuration $\bPhi_t$ at sub-slot $t$ effectively imposes a directional spatial filter, selectively enhancing signal components originating from the beamspace direction $(\theta_p, \phi_q)$.
After $T_1$ sub-slots, the resulting measurement vector $\mathbf{f} = [f_1, \ldots, f_{T_1}]$ exhibits a pronounced peak only when the true angles lie within the sampled grid.

However, due to the absence of strict orthogonality between far-field and near-field steering vectors, the measurement output exhibits a non-sparse energy distribution, where the true angle manifests as a dominant peak accompanied by sidelobe spreading \cite{10620366}.
To obtain a coarse angle estimate, we apply a filtering operation around the main lobe of the observed energy peak.
By setting a power attenuation threshold of 3 dB, the angular sets for the estimated azimuth can be expressed as
\begin{align} \label{eq:angular_support}
    \hat{\bm{\Theta}} = \bigcup \nolimits_{k=1} [\theta_{\text{peak},k}-\Delta \theta_{k},\theta_{\text{peak},k}+\Delta \theta_{k}],
\end{align} 
where $\Delta \theta_{k}$ satisfies
\begin{align}\label{eq:3dB_search}
    \Delta \theta_{k} = \max \left\{ \delta \Big| \mb{f}(\theta_{\text{peak},k} \pm \delta) \ge \frac{\mb{f}(\theta_{\text{peak},k})}{\sqrt{2}}  \right\}.
\end{align}
% Here, the power decay threshold determines the angular resolution of the estimated sets: a larger threshold yields a narrower and more accurate angular interval, while a smaller threshold results in a wider, more conservative estimate.

% \vspace{-5pt}
\subsection{Refined estimation}\label{subsec:refined_es}
The coarse estimation in the previous stage significantly reduces the angular search space, enabling the subsequent super-resolution algorithm to accurately estimate both angles and distances.
This method relies on estimating the covariance matrix of the channel from the sources to the metasurface.
However, since the number of receiver antennas is significantly smaller than the number of metasurface elements, multiple signal stacking is required to obtain a unique estimate of $\mathbf{G} \mathbf{s}$ \cite{ramezani2024efficient}.

Assume that $T_2$ time blocks are used for covariance estimation, and each time block is further divided into $S$ slots ($S \ge \lfloor \frac{N}{N_s} \rfloor$).
The metasurface adopts different phase configurations $\bm{\Phi}_s, s=1,\ldots,S$ at each slot.
By stacking the received signals across slots, we construct the following matrix representation:
\begin{align}
   \tilde{\mathbf{y}}=\tilde{\mathbf{H}} \mathbf{G} \mathbf{s} + \tilde{\mathbf{n}} \in \mathbb{C}^{N_s S \times 1},
\end{align}
% where $\tilde{\mathbf{H}} = [(\mathbf{W}^\H_{\text{BB}}\mathbf{H}\bm{\Phi}_1)^\T,\ldots, (\mathbf{W}^\H_{\text{BB}}\mathbf{H}\bm{\Phi}_S)^\T ]^\T \in \mathbb{C}^{N_s  S \times N}$, 
where $\tilde{\mathbf{H}} = [(\mathbf{H}\bm{\Phi}_1)^\T,\ldots, (\mathbf{H}\bm{\Phi}_S)^\T ]^\T \in \mathbb{C}^{N_s  S \times N}$, $\mathbf{s}$ is defined in \Cref{theo:MatrixModel} 
and $\tilde{\mathbf{n}} = [\mathbf{n}_1^\T,\ldots,\mathbf{n}_S^\T]^\T \in \mathbb{C}^{N_s S\times 1}$ denote the effective stacked channel and noise matrices, respectively.
Following the procedure in \cite{ramezani2024efficient}, the vector $\mathbf{G} \mathbf{s}$ can be estimated via the least squares (LS) method.
Based on this, the sample covariance matrix is computed as
\begin{align} \label{eq:SCM}
    \bm{\Sigma}_s=\frac{1}{T_2}\sum_t \mathbf{G} \mathbf{s}_t ( \mathbf{G} \mathbf{s}_t)^\H=\mathbf{G}\mathbf{S}\mathbf{G}^\H + \sigma^2 (\tilde{\mathbf{H}}^\H \tilde{\mathbf{H}})^{-1},
\end{align}
where $\mathbf{S}=\frac{1}{T_2}\sum_t \mathbf{s}_t\mathbf{s}_t^\H $ denotes the covariance matrix of the transmit symbols.
The eigenvalue decomposition of $\bm{\Sigma}_s$ can be expressed as
\begin{align*}
    \bm{\Sigma}_s = \U_s \bm{\Lambda}_s \U_s^\H + \U_n \bm{\Lambda}_n \U_n^\H,
\end{align*}
where $\U_s \in \mathbb{C}^{N \times K}$ denotes the signal subspace associated with the $K$ largest eigenvalues, and $\U_n \in \mathbb{C}^{N \times (N-K)}$ is its orthogonal complement representing the noise subspace.
The matrices $\bm{\Lambda}_s \in \mathbb{C}^{K \times K}$ and $\bm{\Lambda}_n \in \mathbb{C}^{(N-K) \times (N-K)}$ are diagonal, containing the corresponding eigenvalues of $\bm{\Sigma}_s$.

\begin{figure}[t] 
    \centering
    \includegraphics[width=0.8\linewidth]{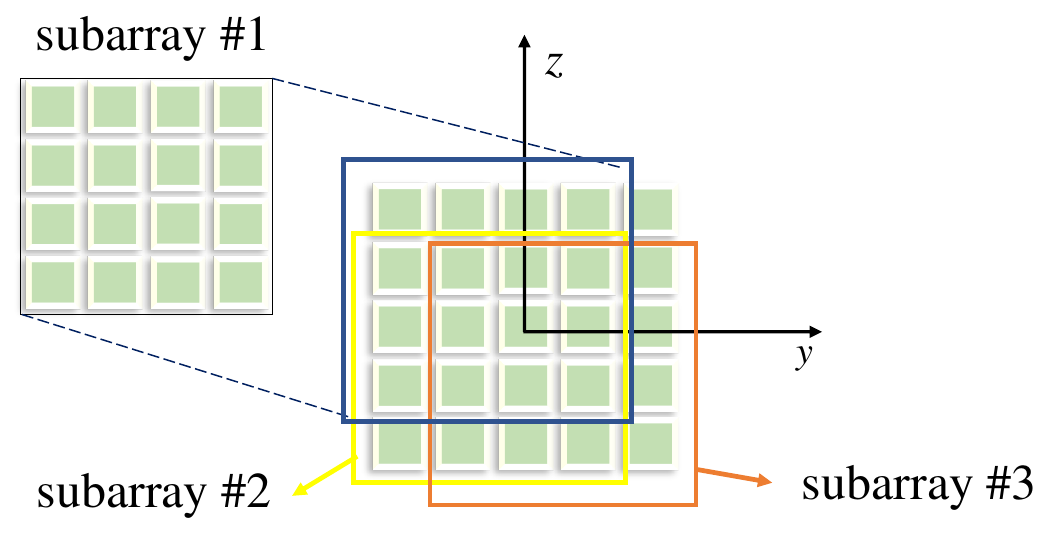}
    \caption{Spatial subarray partitioning of the metasurface. 
    % The first subarray is placed in the top-left region and the second subarray is positioned in the lower-left region, with the third subarray shifted by $1$ unit along the y-axis in comparison to the second. 
    } \label{fig:subarray}
    \vspace{-8pt}
\end{figure}

We partition the metasurface into three subarrays symmetrically distributed with respect to the coordinate origin.
Subarray \#1 comprises the first $2N_h$ elements along the horizontal (y-axis) and the last $2N_v$ elements along the vertical (z-axis), with coordinate ranges $(-N_h,N_h-1)$ for the y-axis and $(-N_v+1,N_v)$ for the z-axis.
Subarray \#2 is obtained by shifting subarray \#1 downward by one unit along the z-axis, resulting in a coordinate range of $(-N_v,N_v-1)$ in the vertical direction.
Similarly, subarray \#3 follows the same structure, as shown in ~\ref{fig:subarray}.
Accordingly, the channel matrix corresponding to the $i$-th subarray has the following structure
\begin{align*}
    \mathbf{G}_i=[\frac{\mathcal{B}_1}{r_1} \bm{\alpha}_{N_i}(\theta_1,\phi_1,r_1),\ldots,\frac{\mathcal{B}_K}{r_K} \bm{\alpha}_{N_i}(\theta_K,\phi_K,r_K)] ,
\end{align*} 
where $\bm{\alpha}_{N_i}(\theta,\phi,r)$ denotes the subarray steering vector obtained by extracting a subset of rows from the full array steering vector $\bm{\alpha}(\theta,\phi,r)$.
That is, the matrix $\mathbf{G}_i$ is obtained by selecting specific rows from the full channel matrix $\mathbf{G}$ while preserving the original column ordering:
\begin{align} \label{eq:Ai=JiA}
\mathbf{G}_i = \mathbf{J}_i \mathbf{G}  \in \mathbb{C}^{(2N_h \times 2N_v) \times K}, \quad i = 1,2,3,
\end{align}
where $\mathbf{J}_i \in \mathbb{C}^{(2N_h \times 2N_v) \times N}$ is a binary row-selection matrix defined element-wise as
% Each row of $\mathbf{J}_i$ contains a single entry equal to 1, with all others being zero. Specifically, $\mathbf{J}_i$ is
\begin{align} \label{eq:define_J}
    [\mb{J}_i]_{m,n}=\begin{cases}
        1, \quad n=m+\lceil \frac{m}{2N_v} \rceil + \varrho_i  \\
        0,\quad \text{else}
    \end{cases}.
\end{align}
% where $\lceil \cdot \rceil$ represents the ceiling function. 
For different values of $i$, we have $\varrho_1 = 0$, $\varrho_2 =-1$ and $\varrho_3 =2N_v$.
This construction ensures that $\mathbf{J}_i$ acts as a structured selector that extracts specific $(2N_h \times 2N_v)$ rows from the $((2N_h+1) \times (2N_v+1))$ rows of $\mb{G}$.

Define the exchange matrix $\mathbf{J}$ as a square matrix with ones on the anti-diagonal and zeros elsewhere.
Based on the geometric symmetry between subarray \#1 and subarray \#3, the following relationship holds
\begin{align*}
    \mb{J}\bm{\alpha}_{N_1}( \gamma_k^a, \gamma_k^e, \beta_k^a, \beta_k^e, \alpha_k) =  \mb{D}( \gamma_k^a, \gamma_k^e) \bm{\alpha}_{N_3}( \gamma_k^a, \gamma_k^e, \beta_k^a, \beta_k^e, \alpha_k),
\end{align*}
where 
\begin{align}\label{eq:D-kronecker}
    \mb{D}(\gamma_k^a&,\gamma_k^e) = \diag(e^{j2(N_h-1)\gamma_k^a},e^{j2(N_h-2)\gamma_k^a},\!\ldots\!,e^{j2(-N_h)\gamma_k^a}) \notag\\
    &\otimes \diag(e^{j2N_v\gamma_k^e},e^{j2(N_v-1)\gamma_k^e},\!\ldots\!,e^{j2(-N_v+1)\gamma_k^e}).
\end{align}
Accordingly, the channel matrices satisfy the following symmetry
\begin{align}\label{eq:symmetry-relationship}
    \mb{J}\mb{G}_1 = &[\mb{D}(\!\gamma_1^a,\!\gamma_1^e) \frac{\mathcal{B}_1}{r_1} \bm{\alpha}_{N_3}(\!\gamma_1^a,\!\gamma_1^e,\!\beta_1^a,\!\beta_1^e,\!\alpha_1),\ldots,\mb{D}(\!\gamma_K^a,\!\gamma_K^e) \notag\\
    &\times \frac{\mathcal{B}_K}{r_K}\bm{\alpha}_{N_3}(\!\gamma_K^a,\!\gamma_K^e,\!\beta_K^a,\!\beta_K^e,\!\alpha_K)].
\end{align}
Similarly, we apply the selection matrices $\mathbf{J}_i$ to extract the corresponding portions of the signal subspace
\begin{align*}
    \U_i= \mb{J}_i \U_s, \quad i=1,2,3,
\end{align*}
where the signal subspace 
% $\mathbf{U}_s$ denotes  spanned by the full channel matrix $\mathbf{G}$, i.e., 
$\mathbf{U}_s = \mathbf{G} \mathbf{T}$ for some full-rank matrix $\mathbf{T} \in \mathbb{C}^{K \times K}$.
We now construct the following spectral function to estimate the azimuth and elevation angles:
\begin{align} \label{eq:angle_search_func}
    f(\gamma^a,\gamma^e) = \det( \mb{W} \J \U_1-\mb{W}\mathbf{D}(\gamma^a,\gamma^e)\U_3)^{-1},
\end{align}
where $\mathbf{W} \in \mathbb{C}^{K \times (2N_h \times 2N_v)}$ is an arbitrary full-rank weighting matrix, and $\mathbf{D}(\gamma^a, \gamma^e)$ is a parameterized diagonal matrix following the Kronecker structure defined in \eqref{eq:D-kronecker}.

It can be observed that when the estimated angular pair $(\gamma^a, \gamma^e)$ coincides with the true values $(\gamma^a_k, \gamma^e_k)$, the $k$-th column of the matrix $\mathbf{J}\mathbf{G}_1 - \mathbf{D}(\gamma^a, \gamma^e)\mathbf{G}_3$ becomes zero, as a direct consequence of the subarray symmetry described in \eqref{eq:symmetry-relationship}.
This leads to a rank deficiency in the matrix $\mathbf{W} \mathbf{J} \mathbf{U}_1 - \mathbf{W}\mathbf{D}(\gamma^a, \gamma^e) \mathbf{U}_3$.
As a result, the directions of DoAs can be estimated by performing a two-dimensional search for the peak of the spectral function $f(\gamma^a, \gamma^e)$ defined in \eqref{eq:angle_search_func}. 
The estimated azimuth and elevation angles are then given by
\begin{align*}\label{eq:spectral-function}
    \!\!(\hat\theta,\hat\phi)\!=\! \Big\{\!\!\arcsin(\frac{\hat{\gamma}^a}{k_c d})\!,\arcsin(\frac{\hat{\gamma}^e}{k_c d}) | (\hat{\gamma}^a\!,\!\hat{\gamma}^e) \!\in\! \arg \!\!\max_{\{\hat{\bm{\Theta}},\hat{\bm{\Psi}}\}} \! f(\gamma^a\!,\!\gamma^e) \! \Big\},
\end{align*}
where $\hat{\bm{\Theta}}$ and $\hat{\bm{\Psi}}$ represent the angular sets regions for azimuth and elevation, respectively, which are obtained via the coarse estimation procedure detailed in \Cref{subsec:beam_filter}.

Having estimated the source's angular parameters, we now proceed to describe the method for distance estimation.
Similar to the angular case, we derive a rotational relationship between the steering vectors associated with subarray \#1 and subarray \#2, given by
\begin{align*}
    \bm{\alpha}_{N_1}(\!\gamma_k^a,\!\gamma_k^e,\!\beta_k^a,\!\beta_k^e,\!\alpha_k)\!=\!e^{j \gamma_k^e} \mb{E}(\!\beta_k^e,\!\alpha_k) \bm{\alpha}_{N_2}(\!\gamma_k^a,\!\gamma_k^e,\!\beta_k^a,\!\beta_k^e,\!\alpha_k),
\end{align*}
where the diagonal matrix $\mathbf{E}(\beta_k^e, \alpha_k)$ is defined as
\begin{align*}
    \mb{E}(\beta_k^e,&\alpha_k) =\diag(e^{j(-N_h)\alpha_k},e^{j(-N_h+1)\alpha_k},\ldots,e^{j(N_h-1)\alpha_k}) \notag \\
    &\otimes \diag(e^{j(-2N_v+1)\beta_k^e},e^{j(-2N_v+3)\beta_k^e},\ldots,e^{j(2N_v-1)\beta_k^e}).
\end{align*}
Based on this relationship, we define a spectral function for distance estimation as
\begin{align} 
  g(\beta^e,\alpha)=\text{det}(\mathbf{W}\mathbf{U}_1-\mathbf{W}\mathbf{E}(\beta^e,\alpha) \mathbf{U}_2)^{-1}.
\end{align}
Given the estimated angles $(\hat{\theta}_k, \hat{\phi}_k)$ from \eqref{eq:angle_search_func}, we perform a grid search over candidate distances and compute the associated $(\beta_k^e, \alpha_k)$ to evaluate the above function. The matrix $\mathbf{W} \mathbf{U}_1 - \mathbf{W} \mathbf{E}(\beta^e, \alpha) \mathbf{U}_2$ becomes rank-deficient when $(\beta^e, \alpha)$ matches the true values, thereby indicating the correct distance.
Accordingly, the estimated distance $\hat{r}_k$ corresponding to the $k$-th signal component is obtained by
\begin{align} \label{eq:r_search_func}
\hat{r}_k \in \arg \max_r g(\beta_k^e, \alpha_k),
\end{align}
where the parameters $\beta_k^e$ and $\alpha_k$ are functions of $r$ and the estimated angles, given by:
\begin{align*}
\beta_k^e = k_c d^2 \cos^2 \hat{\phi}_k/(2 r),
\alpha_k = -k_c d^2 \sin \hat{\theta}_k \sin \hat{\phi}_k/r.
\end{align*}
When $\hat{r}_k > 2(D_h^2 + D_v^2)/\lambda$, i.e., beyond the Fresnel distance threshold defined in \cite{10616028}, the corresponding path can be classified as a far-field signal.
Accordingly, the algorithm simultaneously estimates the positions and numbers of both near-field and far-field sources with a sampling budget of $(T_1 + T_2 \times S)$ points.
% The complete channel estimation procedure is summarized in \Cref{alg:TraESPRIT}.
% The total number of metasurface sampling points required for channel estimation is given by $(T_1 + T_2 \times S)$.

\section{Spatial resolution of the MELA Systems}\label{sec:sp_resolution}

In this section, we compare the MELA system with the traditional ELAA system in terms of spatial resolution.
The HPBW quantifies the angular width of the main lobe at half of its peak power, serving as a measure of beam concentration and an important indicator of the system's sensing capability.
To derive the HPBW expression for the MELA architecture, we model the MELA as a unified transmitter.
Accordingly, the previously designated $M$ receiving antennas now operate in transmission mode to illuminate the metasurface.

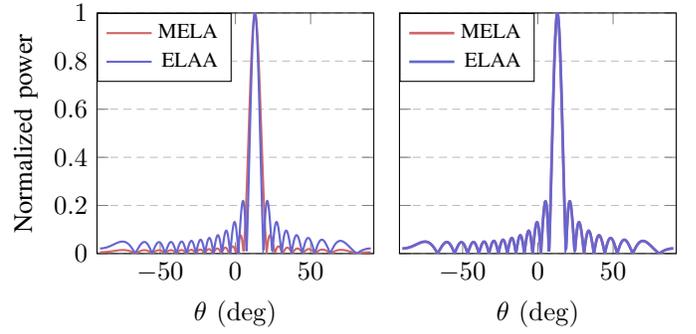
\begin{figure}[t]
    \centering
% \hspace{-8pt} 
\begin{minipage}[t]{0.54\linewidth}
\centering
\begin{tikzpicture}
    \hspace{-4pt} \renewcommand{\axisdefaulttryminticks}{4} 
    \pgfplotsset{every major grid/.style={densely dashed}}       
    \tikzstyle{every axis y label}+=[xshift=4pt] 
    \tikzstyle{every axis x label}+=[yshift=5pt]
    \pgfplotsset{every axis legend/.append style={cells={anchor=east},fill=white, at={(0,1)}, anchor=north west, font=\footnotesize}}
    \begin{axis}[
        width=1.1\columnwidth,
        height=1\columnwidth,
        ymajorgrids=true,
        scaled ticks=true,
        xlabel = { $\theta\ (\mathrm{deg})$ },
        xmin=-92, xmax=92,
        ylabel = { Normalized power },
        ylabel style={at={(axis description cs:-0.18,0.5)}}, 
        ymin = 0, ymax = 1,
        scaled ticks=true,
        ]
        \addplot[smooth, RED!60!white, line width=0.5pt ] table[x=theta_range, y=TRMS] {hpbw_2.txt};
        \addlegendentry{MELA};
        \addplot[smooth, BLUE!60!white, line width=0.5pt ] table[x=theta_range, y=ELAA] {hpbw_2.txt};
        \addlegendentry{ELAA};

        \end{axis}
\end{tikzpicture}
\end{minipage}%
% \hspace{-1pt} 
\begin{minipage}[t]{0.54\linewidth}
    \centering
    \begin{tikzpicture}
        \centering
        \hspace{-10pt} \renewcommand{\axisdefaulttryminticks}{4} 
        \pgfplotsset{every major grid/.style={densely dashed}}       
        \tikzstyle{every axis y label}+=[yshift=-10pt] 
        \tikzstyle{every axis x label}+=[yshift=5pt]
        \pgfplotsset{every axis legend/.append style={cells={anchor=east},fill=white, at={(0,1),sacel=0.2}, anchor=north west, font=\footnotesize}} %nodes={scale=0.2, transform shape}}
        \begin{axis}[
            width=1.1\columnwidth, height=1\columnwidth, ymajorgrids=true, scaled ticks=true, xlabel = { $\theta\ (\mathrm{deg})$  }, ylabel = { }, scaled ticks=true,xmin=-92, xmax=92,ymin = 0, ymax = 1,yticklabel=\empty,
            ] 
            \addplot[smooth, RED!60!white, line width=0.5pt ] table[x=theta_range, y=TRMS] {hpbw_1.txt};
            \addlegendentry{MELA};
            \addplot[smooth, BLUE!60!white, line width=0.5pt ] table[x=theta_range, y=TRMS] {hpbw_1.txt};
            \addlegendentry{ELAA};
            \end{axis}
    \end{tikzpicture}
    \end{minipage} 
\caption{ Normalized array-factor power versus azimuth angle. Results in red are computed using \eqref{eq:AF}, whereas those in blue are obtained from the array-factor expression in Eq. (16) of \cite{xiong2024design}.
\textbf{Left:} Three near-field feeds located at distances of $[0.04, 0.05, 0.06]m$; HPBWs of MELA and ELAA are $9.02^\circ$ and $6.85^\circ$, respectively.
\textbf{Right:} Three far-field feeds located at distances of $[2.5, 2.6, 2.7]m$; HPBWs are $7.20^\circ$ (MELA) and $6.85^\circ$ (ELAA).
} 
\label{fig:HPBW}
\vspace{-5pt}
\end{figure}

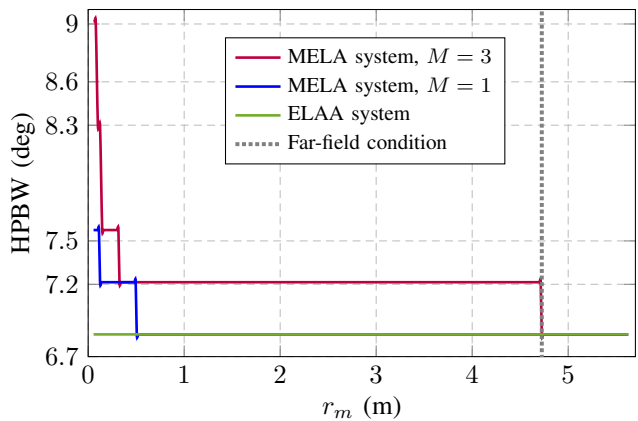
\begin{figure}[t]
    \centering
    \hspace{-0.6cm} 
\begin{minipage}[t]{1\linewidth}
\centering
\begin{tikzpicture}
    \renewcommand{\axisdefaulttryminticks}{4} 
    \pgfplotsset{every major grid/.style={densely dashed}}       
    \tikzstyle{every axis y label}+=[yshift=-10pt] 
    \tikzstyle{every axis x label}+=[yshift=5pt]
    \pgfplotsset{every axis legend/.append style={cells={anchor=west},legend columns=1,fill=white, at={(0.25,0.92)}, anchor=north west, font=\footnotesize}}
    \begin{axis}[
        width=1\columnwidth,
        height=0.7\columnwidth,
        ymajorgrids=true,
        scaled ticks=true,
        xlabel = {$r_m$ (m)},
        xmin=0, xmax=5.7,
        x label style={yshift=2pt},
        ylabel = {HPBW (deg)},
        ylabel style={yshift=-3pt},
        ymin = 6.7, ymax =9.1,
        ytick={6.7,7.2,7.5,8.3,8.6,9.0},
        % ymode = log,
        ytick distance=2,
        grid=both,
        x tick scale label style={
        font=\small,
        xshift=-5pt,
        yshift=-2pt    
        },
        ]
        \addplot[smooth, purple, line width=1pt ] table[x=dd, y=mela1] {hpbw.txt};
        \addlegendentry{MELA system, $M=3$};
        \addplot[smooth, blue, line width=1pt ] table[x=dd, y=mela3]{hpbw.txt};
        \addlegendentry{MELA system, $M=1$};
        \addplot[smooth, green1, line width=1pt] table[x=dd, y=elaa] {hpbw.txt};
        \addlegendentry{ELAA system};
        \addplot[densely dotted, color=gray, line width=1.5pt] coordinates { (4.725,6) (4.725,9.1) };
        \addlegendentry{Far-field condition};
        \end{axis}
\end{tikzpicture}
\end{minipage}%
\caption{ The HPBW versus the distance $r_m$. The feed antennas is located at a random angle, with the distance set as $r_m = \sqrt{D_h^2 + D_v^2} [0.5{:}0.5{:}50]$.
} 
\label{fig:HPBW with d}
\vspace{-0.4cm}
\end{figure}

For the passive metasurface, the resulting radiation field depends on both the incident field and the applied phase shifts.
based on the symmetry of the overall channel, the radiated power observed in the far-field outgoing direction $(\theta_r,\phi_r)$ can be expressed as \cite{balanis2016antenna}
\begin{align} \label{eq:y_in_HPBW}
    y = \textstyle \sum_{m=1}^M \bm{\alpha}^\H(\theta_r,\phi_r) \bPhi \mathbf{h}(\theta_m,\phi_m,r_m),
\end{align}
where $\mathbf{h}(\theta_m,\phi_m,r_m) = [\mathbf{H}^\H]_{(:,m)}$ corresponds to the $m$-th column of $\mathbf{H}^\H$ and represents the channel between the $m$-th feed antenna and the metasurface.
Here, the metasurface uses a phase compensation strategy that maximizes power transmission.
% where $\bm{g}(\theta_r,\phi_r,\infty)=\frac{\mathcal{B}_r}{r_r}\bm{\alpha}(\theta_r,\phi_r)$ denotes the far-field channel vector with $r_r$ representing the distance in the outgoing direction and $\mathcal{B}_r$ defined in \Cref{theo:main}.
% where $\bm{\Omega} = \mathrm{diag}(\mathbf{w})$ is a diagonal weighting matrix with $\mathbf{w} = [e^{j \omega_1}, e^{j \omega_2}, \dots, e^{j \omega_{N}}]^\T$ denoting the phase control vector.
Let
\begin{align*}
    \mathbf{c}(\theta_r,\phi_r,\theta_m,\phi_m,r_m) = \textstyle \sum_{m=1}^M \bm{\alpha}(\theta_r,\phi_r) \odot  \mathbf{h}(\theta_m,\phi_m,r_m),
\end{align*}
then \eqref{eq:y_in_HPBW} can be simplified as $y = \mathbf{c}^\H(\theta_r,\phi_r,\theta_m,\phi_m,r_m) \bm{\omega}$,
which is maximized when the phase of metasurface compensates for that of the vector $\mathbf{c}$, i.e.,
\begin{align}
\bm{\omega} = e^{j \arg(\mathbf{c}(\theta_r,\phi_r,\theta_m,\phi_m,r_m))}.
\end{align}
Accordingly, for a given observation direction $(\theta_{\text{out}},\phi_{\text{out}})$, the array factor is computed as
\begin{equation*}\label{eq:AF}
    \begin{split}
    & \text{AF}(\theta_{\text{out}},\phi_{\text{out}})= \mathbf{c}^\H(\theta_{\text{out}},\phi_{\text{out}},\theta_m,\phi_m,r_m) \bm{\omega} \\
    &= \bm{\alpha}(\theta_{\text{out}},\phi_{\text{out}}) \Big( \bm{\alpha}^\H(\theta_r,\phi_r)  \odot \text{mag}\big(\sum_{m=1}^M \mathbf{h}(\theta_m,\phi_m,r_m) \big) \Big) .
    \end{split}
\end{equation*} 
Half-power beamwidth can be found by determining the half-power point $(\theta_{\text{out}},\phi_{\text{out}})$ that satisfies
\begin{align} \label{eq:hpbw}
    20 \log(\frac{\text{AF}(\theta_{\text{out}},\phi_{\text{out}})}{AF_{\text{max}}})=-3.
\end{align}
As implied by $\text{AF}(\theta_{\text{out}},\phi_{\text{out}})$, when the feed antennas are placed close to the metasurface, the HPBW expression in \eqref{eq:hpbw} becomes explicitly dependent on the feed-to-metasurface distance.
In this regime, the per-element weights induced by $\text{mag}\big(\sum_{m=1}^M \mathbf{h}(\theta_m,\phi_m,r_m) \big)$ vary significantly across the aperture, resulting in a broader HPBW compared to conventional ELAA systems, as illustrated in Fig.~\ref{fig:HPBW} and Fig.~\ref{fig:HPBW with d}.

In the asymptotic regime where the feeds are sufficiently far from the transmissive array, or, equivalently, when the per-element magnitudes in $\text{mag}\big(\sum_{m=1}^M \mathbf{h}(\theta_m,\phi_m,r_m) \big)$ are approximately identical (e.g. $M=1$), the aperture is uniformly illuminated.
In this regime, the metasurface phase shifts simply compensate the incident phases, and
\begin{align*} 
    \frac{\text{AF}(\theta_{\text{out}},\phi_{\text{out}})}{AF_{\text{max}}}\approx \frac{\sin \Psi_1}{\Psi_1} \cdot \frac{\sin \Psi_2}{\Psi_2}, 
\end{align*} 
where $\Psi_1 = k_c d (\sin\theta_r - \sin\theta_{\text{out}})$ and $\Psi_2 = k_c d (\sin\phi_r - \sin\phi_{\text{out}})$. 
The resulting HPBW expression coincides with that of a conventional antenna array \cite{xiong2024design}.

% setting中没有写RIS的参数设置

\section{Numerical Results}
\label{sec:simulation}
In this section, we present numerical simulations to validate the proposed model and algorithms.
Consider a metasurface consisting of $21 \times 21$ units, with all elements spaced at half-wavelength intervals.
The system operates at a frequency of 28 GHz.
Both the metasurface units and the receiving antenna elements are assumed to have sizes of $d_r = d_t = \lambda/2$.
In \Cref{sec:num_result_cst}, we validate the proposed electric-field model via full-wave electromagnetic simulations.
Subsequently, in \Cref{sec:num_result_1}, we analyze the impact of various system parameters on the proposed electromagnetic channel model.
Finally, in \Cref{sec:num_result_2}, we validate the effectiveness of the proposed beamspace filtering method and compare our channel estimation algorithm with two other hybrid-domain estimation techniques and the Cramér-Rao bound (CRB).

\subsection{Full-wave Simulation} \label{sec:num_result_cst}
\begin{figure}[t]
    \centering
    \begin{subfigure}{0.78\linewidth}
        \centering
        \includegraphics[width=\linewidth]{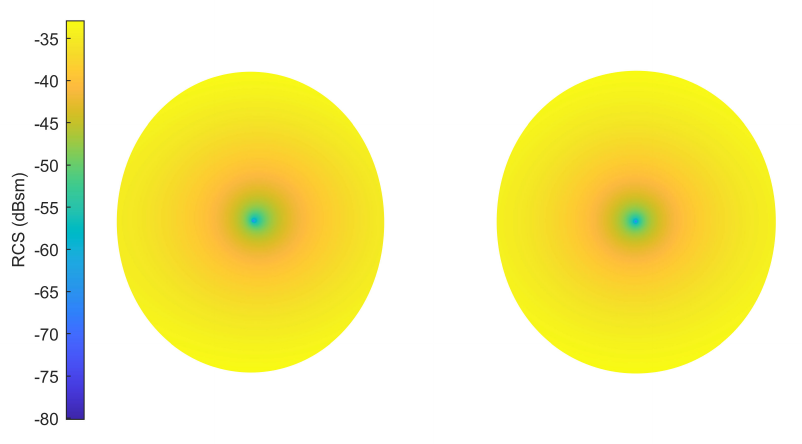}
    \end{subfigure}
    \vspace{6pt}
    \begin{subfigure}{0.8\linewidth}
        \centering
        \includegraphics[width=\linewidth]{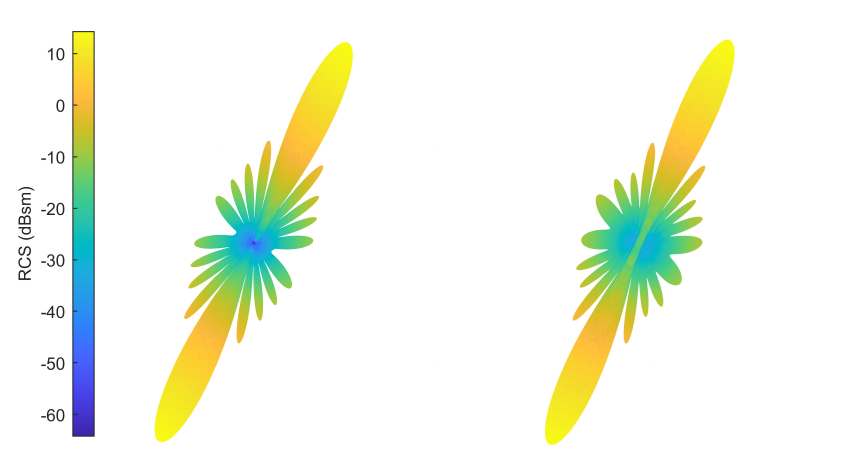}
    \end{subfigure}

    \caption{
    \textbf{Top:} Single-element 3D RCS lobes. Left: analytical model; right: CST simulation (copolarized), \(\phi=0^\circ\) cut. 
    \textbf{Bottom:} Array-level 3D RCS lobes. Left: analytical model; right: CST simulation (copolarized), \(\phi=0^\circ\) cut.
    The source is located on the x-axis under normal incidence, and the reception is co-polarized.}
    \label{fig:rcs_lobes}
    \vspace{-8pt}
\end{figure}

We first validate the proposed electromagnetic model against full-wave simulations in CST Studio Suite.
We analyze the monostatic Radar Cross Section (RCS) under a plane wave normally incident along the $x$-axis and co-polarized reception.
Based on the formulation in \eqref{eq:theo1}, we obtain the single-element RCS by taking a far-field observation and a vanishing receive aperture:
\begin{align*}
    \text{RCS}^{\text{element}} = \lim_{\|\mathbf{r}-\mathbf{t}\| \to \infty} 4 \pi \|\mathbf{r}-\mathbf{t}\|^2 \|\bar{\mathbf{E}}_r\|^2/\|\mathbf{E}^i\|^2.
    % &= \frac{(1+\Gamma)^2k_c^2 d_t^4 }{4 \pi}  \sinc^2\!\big( \frac{k_c u_y d_t}{2}  \big) \!\sinc^2\!\big(  \frac{k_c u_z d_t}{2}  \big) \! \cos^2 \phi_{k}.
\end{align*}
For an $N$-element metasurface, the array RCS follows by coherent superposition of the element responses:
\begin{align*}
    &\text{RCS}^{\text{element}} = \lim_{\|\mathbf{r}-\mathbf{t}\| \to \infty} 4 \pi  \lVert \sum_{n=1}^{N} \|\mathbf{r}-\mathbf{t}_n\|^2 \bar{\mathbf{E}}_r(\mathbf{t}_n)\rVert ^2/\|\mathbf{E}^i\|^2.
\end{align*}
The results in Fig.~\ref{fig:rcs_lobes} show excellent agreement between the analytical model and CST: both the 3D RCS lobes of a single element and those of the full metasurface exhibit matching main-lobe shapes and sidelobe trends, confirming the accuracy of the proposed model.
 
\subsection{Impact of System Parameters on the Channel} \label{sec:num_result_1}
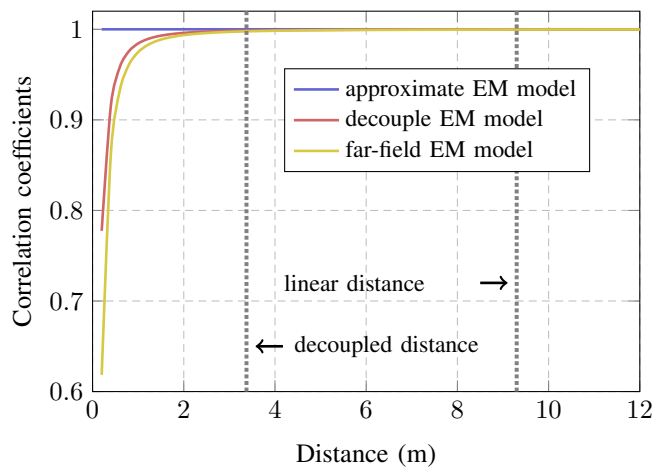
\begin{figure}[t]
    \centering
    % \hspace{-0.5cm} 
\begin{minipage}[t]{1\linewidth}
\centering
\begin{tikzpicture}
    \renewcommand{\axisdefaulttryminticks}{4} 
    \pgfplotsset{every major grid/.style={densely dashed}}       
    \tikzstyle{every axis y label}+=[yshift=-10pt] 
    \tikzstyle{every axis x label}+=[yshift=5pt]
    \pgfplotsset{every axis legend/.append style={cells={anchor=west},fill=white, at={(0.35,0.85)}, anchor=north west, font=\small}}
    \begin{axis}[
        width=0.95\columnwidth,
        height=0.7\columnwidth,
        ymajorgrids=true,
        scaled ticks=true,
        xmin=0, xmax=12,
        xlabel = {Distance (m)},
        x label style={yshift=-3pt},
        ylabel = { Correlation coefficients },
        ymin = 0.6, ymax = 1.02,
        % ytick distance=2,
        grid=both,
        x tick scale label style={
        font=\small,
        },
        ]
        \addplot[smooth, BLUE!60!white, line width=1pt ] table[x=dd_iter, y=app] {beam_fig1.txt};
        \addlegendentry{approximate EM model};
        \addplot[smooth, RED!60!white, line width=1pt] table[x=dd_iter, y=de] {beam_fig1.txt};
        \addlegendentry{decouple EM model};
        \addplot[smooth, yellow!80!black , line width=1pt ] table[x=dd_iter, y=far] {beam_fig1.txt};
        \addlegendentry{far-field EM model};
        \addplot[ densely dotted, color=gray, line width=1.5pt] coordinates { (3.375,0.6) (3.375,1.1) };
        \draw[<-,line width = 1pt] (axis cs:3.375+0.2,0.65) --  (axis cs:3.375+0.8,0.65) ;
        \node[anchor=west] at (axis cs:3.375+0.85,0.65) {\textcolor{black} {\small decoupled distance}};
        \addplot[ densely dotted, color=gray, line width=1.5pt] coordinates { (9.30,0.6) (9.30,1.1) };
        \draw[->,line width = 1pt] (axis cs:9.30-0.8,0.72) --  (axis cs:9.30-0.2,0.72) ;
        \node[anchor=west] at (axis cs:4.5,0.72) {\textcolor{black} {\small linear distance}};
        \end{axis}
\end{tikzpicture}
\end{minipage}
\caption{ The correlation coefficients versus the distance between the geometric center of metasurface and receiver.
} 
\label{fig:beam_power}
\hspace{-10pt} 
\end{figure}

Consider a scenario where three single-dipole sources transmit uplink signals, and the receiving antennas employ a uniform linear array consisting of 15 antennas spaced at half-wavelength intervals.
We define the normalized correlation coefficient based on the received signals as $\rho_{(\cdot)} =  |  \mathbf{y}^\H_{(\cdot)} \mathbf{y}_{\text{true}}  |^2 / (| \mathbf{y}_{(\cdot)}|^2  | \mathbf{y}_{\text{true}}|^2 )$, which serves to quantify the correlation between different approximate models and the accurate electromagnetic model.
% \[
% \vspace{-1pt}
%     \rho_{(\cdot)} =  | \langle \mathbf{y}^\H_{(\cdot)} \mathbf{y}_{\text{true}} \rangle |^2 / (| \mathbf{y}_{(\cdot)}|^2  | \mathbf{y}_{\text{true}}|^2 ).
%     \vspace{-1pt}
% \]
Fig.~\ref{fig:beam_power} illustrates the correlation coefficients between the true channel responses and three analytical models—the approximate model \eqref{eq:channel model}, the decoupled model \eqref{eq:H_decoupled}, and the linear model \eqref{eq:H_far}—under different placement distances.
It is observed that the approximate electromagnetic model closely matches the true field, serving as an upper bound in terms of the correlation coefficient.
When the metasurface-receiver distance exceeds the decoupling distance, the decoupled model also exhibits strong agreement with the true model, as indicated by a beam power value approaching 1.
Furthermore, once the distance surpasses the linear distance, the beam power curves of the linear, decoupled, and electromagnetic models converge, demonstrating their consistency in the far-field regime.

\begin{figure}[t!] 
\hspace{-5pt} 
    \centering
    \begin{subfigure}[t]{1\linewidth}
    \begin{minipage}[t]{1\linewidth}
    \centering
    \begin{tikzpicture}
        % \hspace{-0.5cm} 
    \begin{axis}[
        width=0.9\columnwidth,
        height=0.5\columnwidth,
        ybar,
        xlabel={Number of eigenvalues},
        ylabel={Normalized strength},
        xmin=1, xmax=15.5,
        ymin=0, ymax=1,
        xtick={1,2,...,15},
        ytick={0,0.5,1},
        xtick=data,
        grid=both,
        grid style={gray!30,very thin},
        legend style={cells={anchor=west},
        at={(0.95,0.95)},
        anchor=north east,
        font=\small,
        fill=white,
        draw=black
        },
        ybar interval=0.8pt,
    ]
        \addplot+[blue,  fill=blue!60]   table[x=num,y=500d]  {DOF1.txt}; \addlegendentry{$\| d_c\|=500d$}
        \addplot+[orange,fill=orange!60,bar shift=3.5pt] table[x=num,y=100d]  {DOF1.txt}; \addlegendentry{$\| d_c\|=100d$}
        \addplot+[yellow!80!black,fill=yellow!60,bar shift=6.5pt] table[x=num,y=10d] {DOF1.txt}; \addlegendentry{$\| d_c\|=10d$}
        \addplot+[purple,fill=purple!60,bar shift=9.5pt] table[x=num,y=5d]  {DOF1.txt}; \addlegendentry{$\| d_c\|=5d$}

    \end{axis}
    \end{tikzpicture}
    \end{minipage}
% \caption{ The eigenvalues of $\mathbf{H}$ in decreasing order with element size $d_r=d_t=\lambda/2$.
% } 
    \end{subfigure}

    \begin{subfigure}[t]{1\linewidth}
        \centering
    % \hspace{-0.5cm} 
    \begin{minipage}[t]{1\linewidth}
    \centering
    \begin{tikzpicture}
        % \hspace{-0.5cm} 
    \begin{axis}[
        width=0.9\columnwidth,
        height=0.5\columnwidth,
        ybar,
        xlabel={Number of eigenvalues},
        ylabel={Normalized strength},
        xmin=1, xmax=15.5,
        ymin=0, ymax=0.5,
        xtick={1,2,...,15},
        ytick={0,0.5,1},
        xtick=data,
        grid=both,
        grid style={gray!30,very thin},
        legend style={cells={anchor=west},
        at={(0.95,0.95)},
        anchor=north east,
        font=\small,
        fill=white,
        draw=black
        },
        ybar interval=0.8pt,
    ]
        \addplot+[blue,  fill=blue!60]   table[x=num,y=1lambda]  {DOF2.txt}; \addlegendentry{$d_t=\lambda$}
        \addplot+[orange,fill=orange!60,bar shift=3.5pt] table[x=num,y=2lambda]  {DOF2.txt}; \addlegendentry{$d_t=2\lambda$}
        \addplot+[yellow!80!black,fill=yellow!60,bar shift=6.5pt] table[x=num,y=5lambda] {DOF2.txt}; \addlegendentry{$d_t=5\lambda$}
        \addplot+[purple,fill=purple!60,bar shift=9.5pt] table[x=num,y=20lambda]  {DOF2.txt}; \addlegendentry{$d_t=20\lambda$}

    \end{axis}
    \end{tikzpicture}
\end{minipage}
% \caption{ The eigenvalues of $\mathbf{H}$ in decreasing order with TRIS-receiver distance $d_c=10\lambda$. $d_r$ is set as $d_r=d_t$.
% } 
\vspace{-10pt}
\end{subfigure}
\caption{ The distribution of the normalized eigenvalue strength under different parameter settings. \textbf{Top:} The eigenvalues of $\mathbf{H}$ in decreasing order with element size $d_r=d_t=\lambda/2$. \textbf{Bottom:} The eigenvalues of $\mathbf{H}$ in decreasing order with TRIS-receiver distance $d_c=10\lambda$. $d_r$ is set as $d_r=d_t$.
} 
\label{fig:DOF}
\end{figure}
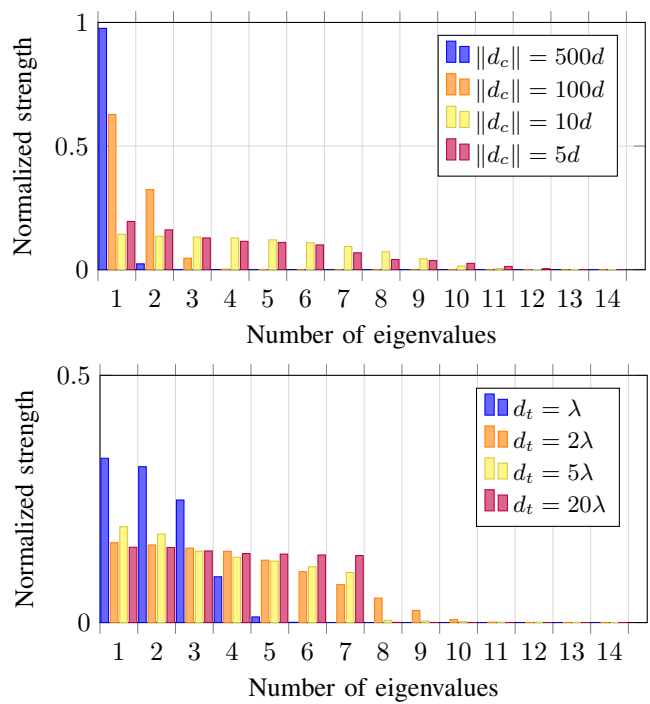

In Fig.~\ref{fig:DOF}, we examine the impact of physical parameters on the degrees of freedom (DoF) of the metasurface-receiver channel in the near-field scenario.
A higher DoF implies that the channel can support a greater number of independent data streams at a given signal-to-noise ratio.
We present in Fig.~\ref{fig:DOF} the variations in the eigenvalue distribution of $\mathbf{H}$ with respect to the metasurface-receiver center distance $d_c$, as well as the element sizes of the transmissive unit cell $d_t$ and the receiving antenna $d_r$.
It is observed that the DoF of the channel decreases as $d_c$ increases, which manifests as rank deficiency in the channel matrix, where a few dominant eigenvalues emerge while the remaining ones diminish.
Meanwhile, increasing the element size and thereby enlarging the antenna aperture leads to a higher DoF.

\vspace{-5pt}
\subsection{Channel Estimation} \label{sec:num_result_2}
To validate the effectiveness of the beamspace filtering method introduced in \Cref{subsec:beam_filter}, a full-space scan is performed over both azimuth and elevation angles ranging from $-90^\circ$ to $90^\circ$.
Three sources are located at $(-59.5^\circ, 21.4^\circ)$, $(44.4^\circ, 8.1^\circ)$, and $(28.7^\circ, -43.3^\circ)$.
The energy attenuation threshold is set to 3 dB, the SNR to $-5$ dB, and all other parameters are kept consistent with those in Fig.~\ref{fig:beam_power}.
The resulting spatial observation function, obtained with a quantization level of $P = Q = 40$, is shown in Fig.~\ref{fig:DDBF}.
As observed, the function exhibits distinct peaks at the true signal locations.
The angular sets coarsely estimated via beamspace filtering correspond to the regions within the 3-dB amplitude attenuation threshold, highlighted in red.

% \begin{figure}[t]
%         \hspace{-8pt}
%     \centering
%     \begin{subfigure}[t]{0.55\linewidth} 
%         \centering    \includegraphics[width=\linewidth]{DDBF.pdf} 
%         % \caption{The quantization numbers are $P=Q=40$.}
%         \label{fig:DDBF_a}
%     \end{subfigure}
%     \hspace{-8pt}
%     \begin{subfigure}[t]{0.5\linewidth}  
%         \centering
%         \includegraphics[width=\linewidth]{DDBF2.pdf}  
%         % \caption{The quantization numbers are $P=Q=10$.}
%         \label{fig:DDBF_b}
%     \end{subfigure}
%     \caption{The distribution of the measurement equation in the angular space.}
%     \label{fig:DDBF}
% \end{figure}
\begin{figure}[t]
    \centering  
        \hspace{-0.2cm}
\includegraphics[width=0.95\linewidth]{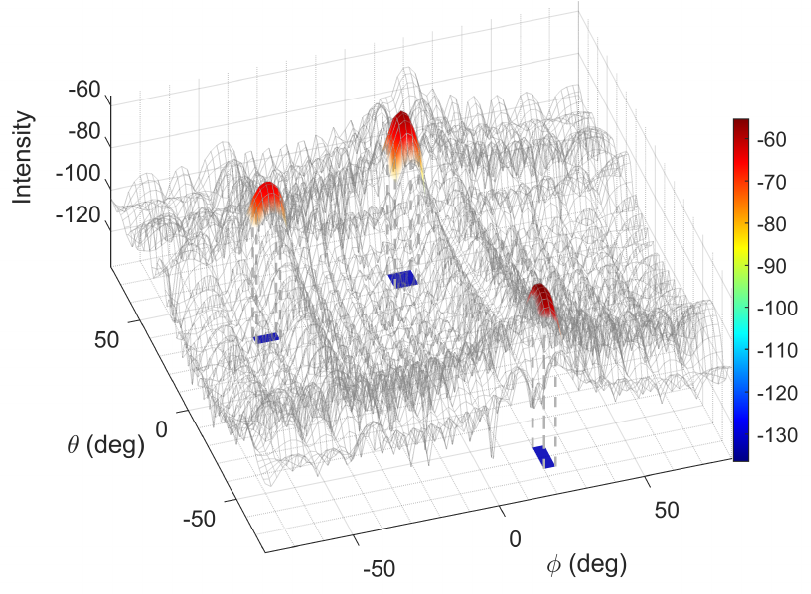} 
    \caption{The distribution of the measurement equation in the angular space.}
    \label{fig:DDBF}
\end{figure}
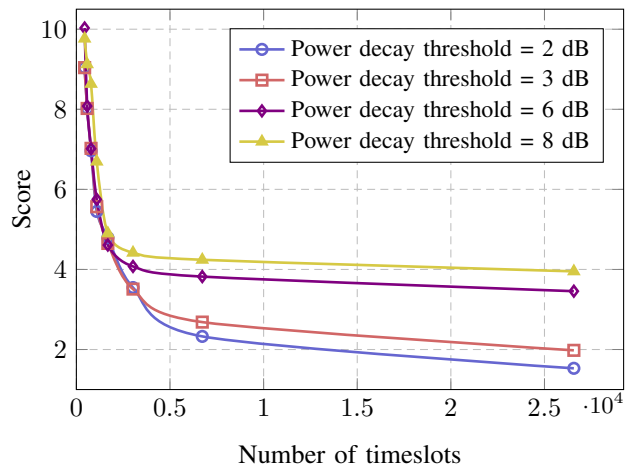
\begin{figure}[t]
    \centering
    \hspace{-0.6cm} 
\begin{minipage}[t]{1\linewidth}
\centering
\begin{tikzpicture}
    \renewcommand{\axisdefaulttryminticks}{4} 
    \pgfplotsset{every major grid/.style={densely dashed}}       
    \tikzstyle{every axis y label}+=[yshift=-10pt] 
    \tikzstyle{every axis x label}+=[yshift=5pt]
    \pgfplotsset{every axis legend/.append style={cells={anchor=east},fill=white, at={(0.28,0.95)}, anchor=north west, font=\small}}
    \begin{axis}[
        width=1\columnwidth,
        height=0.75\columnwidth,
        ymajorgrids=true,
        scaled ticks=true,
        xlabel = { Number of timeslots },
        xmin=0,
        % , xmax=92,
        x label style={yshift=-4pt},
        ylabel = { score (degree) },
        ylabel style={yshift=-3pt},
        ymin = 1, ymax = 10.5,
        ytick distance=2,
        grid=both,
        x tick scale label style={
        at={(rel axis cs:0.9,-0.05)}, 
        anchor=west,            % 左对齐
        xshift=5pt,             % 微调位置
        yshift=2pt,
        font=\small
        },
        ]
        \addplot[smooth, BLUE!60!white, line width=1pt, mark=o ] table[x=time_slot, y=2db] {fig2_data.txt};
        \addlegendentry{Power decay threshold = 2 dB};
        \addplot[smooth, RED!60!white, line width=1pt, mark=square] table[x=time_slot, y=3db] {fig2_data.txt};
        \addlegendentry{Power decay threshold = 3 dB};
        \addplot[smooth, violet, line width=1pt, mark=diamond ] table[x=time_slot, y=6db] {fig2_data.txt};
        \addlegendentry{Power decay threshold = 6 dB};
        \addplot[smooth, yellow!80!black, line width=1pt, mark=triangle*] table[x=time_slot, y=8db] {fig2_data.txt};
        \addlegendentry{Power decay threshold = 8 dB};
        \end{axis}
\end{tikzpicture}
\end{minipage}%
\caption{ The weighted score versus the number of sample spaces under different power decay thresholds.
The SNR is set to -5 dB. Results are averaged over 500 independent trials.
} 
\label{fig:score}
\end{figure}

In Fig.~\ref{fig:score}, we further examine how the power attenuation threshold and the number of scan samples affect the estimation algorithm.
To quantitatively evaluate the estimation accuracy, a weighted score metric is introduced and defined as
\[
\vspace{-1pt}
    \text{score} = \rho \cdot \bar{L} +(1-\rho)\cdot \text{MSE}_{\text{a}},
    \vspace{-1pt}
\]
where $\bar{L}$ denotes the average width of the estimated angular interval, $\rho$ is a weighting parameter that balances the trade-off between interval width and estimation bias, and $\text{MSE}_{\text{a}}$ represents the mean squared error of angle estimation.
It can be observed that a larger power attenuation threshold produces coarser but more stable estimates, thereby increasing the probability that the true angle lies within the estimated interval.
In contrast, a smaller threshold yields narrower intervals that may exclude the true angle, which degrades the performance of subsequent fine channel estimation.
Moreover, increasing the number of scan samples requires more pilot symbols and leads to higher computational complexity.
The results in Fig.~\ref{fig:score} demonstrate that, when the channel remains stable over an extended pilot duration, a power attenuation threshold of 3~dB (corresponding to the inflection point of the performance curve) and a scan range of $[-90^\circ \!:\! 2^\circ \!:\! 90^\circ]$ provide an effective system configuration.
Beyond these settings, further increasing the sample size or reducing the power attenuation threshold yields only marginal performance improvement.
% \Cref{fig:score} also shows that the coarse beamspace filtering method (with specific numerical values omitted) achieves angle estimation errors within $2^\circ$, which is sufficient for applications where moderate estimation accuracy is acceptable, such as in low-precision target detection.

\begin{figure}[t]
    \centering
    % \hspace{-5pt} 
    \begin{subfigure}[t]{1\linewidth}
    \centering
    \begin{tikzpicture}       \renewcommand{\axisdefaulttryminticks}{4} 
        \pgfplotsset{every major grid/.style={densely dashed}}    
        \tikzstyle{every axis y label}+=[yshift=-5pt] 
        \tikzstyle{every axis x label}+=[yshift=-2pt]
        \pgfplotsset{every axis legend/.append style={cells={anchor=west},fill=white, at={(0.01,0.45)}, anchor=north west, font=\small}}
        \begin{axis}[
            width=0.95\columnwidth,
            height=0.62\columnwidth,
            ymajorgrids=true,
            scaled ticks=true,
            xlabel = {SNR},
            ylabel = {MSE (degree)},
            ytick distance=2,
            ymode = log,
            ytick={1e-5,1e-4,1e-3,1e-2,1e-1},
            grid=both,
            xmin  = -16,
            xmax = 16,
            x tick scale label style={
            font=\small,
            xshift=-5pt,
            yshift=-2pt    
            },
            ]
            \addplot[smooth, BLUE!60!white, line width=1pt, mark=o ] table[x=snr, y=ang_T2] {mse_hybrid.txt};
            \addlegendentry{hybrid-field, $T_2=100$};
            \addplot[smooth, RED!60!white, line width=1pt, mark=square] table[x=snr, y=ang_T5] {mse_hybrid.txt};
            \addlegendentry{hybrid-field, $T_2=500$};
            \addplot[smooth, yellow!80!black, line width=1pt, mark=diamond ] table[x=snr, y=ang_T2] {mse_near.txt};
            \addlegendentry{near-field, $T_2=100$};
            \addplot[smooth, violet, line width=1pt, mark=triangle*] table[x=snr, y=ang_T5] {mse_near.txt};
            \addlegendentry{near-field, $T_2=500$};
        \end{axis}
    \end{tikzpicture}
    
    % \caption{Angular-domain performance}
    \end{subfigure}

    \begin{subfigure}[t]{1\linewidth}
    \centering
    \begin{tikzpicture}        \renewcommand{\axisdefaulttryminticks}{4} 
        \pgfplotsset{every major grid/.style={densely dashed}}       
        \tikzstyle{every axis y label}+=[yshift=-5pt] 
        \tikzstyle{every axis x label}+=[yshift=-2pt]
        \pgfplotsset{every axis legend/.append style={cells={anchor=west},fill=white, at={(0.01,0.45)}, anchor=north west, font=\small}}
        \begin{axis}[
            width=0.95\columnwidth,
            height=0.62\columnwidth,
            ymajorgrids=true,
            scaled ticks=true,
            xlabel = {SNR},
            ylabel = {MSE (meter)},
            ytick distance=2,
            ymode = log,
            ytick={1e-1,1e+0,1e+1,1e+2},
            grid=both,
            xmin  = -16,
            xmax = 16,
            x tick scale label style={
            font=\small,
            xshift=-5pt,
            yshift=-2pt    
            },
            ]
            \addplot[smooth, BLUE!60!white, line width=1pt, mark=o ] table[x=snr, y=dis_T2] {mse2_hybrid.txt};
            \addlegendentry{hybrid-field, $T_2=100$};
            \addplot[smooth, RED!60!white, line width=1pt, mark=square] table[x=snr, y=dis_T5] {mse2_hybrid.txt};
            \addlegendentry{hybrid-field, $T_2=500$};
            \addplot[smooth, yellow!80!black, line width=1pt, mark=diamond ] table[x=snr, y=dis_T2] {mse2_near.txt};
            \addlegendentry{near-field, $T_2=100$};
            \addplot[smooth, violet, line width=1pt, mark=triangle*] table[x=snr, y=dis_T5] {mse2_near.txt};
            \addlegendentry{near-field, $T_2=500$};
        \end{axis}
    \end{tikzpicture}
    % \caption{Distance-domain performance}
    \end{subfigure}
    \caption{The mean squared error of our proposed algorithm varies with the SNR under different number of time slots. \textbf{Top:} MSE in the angular domain. \textbf{Bottom:} MSE in the distance domain.
    }
    \label{fig:MSE_T_SNR}
    \vspace{-10pt}
    
\end{figure}
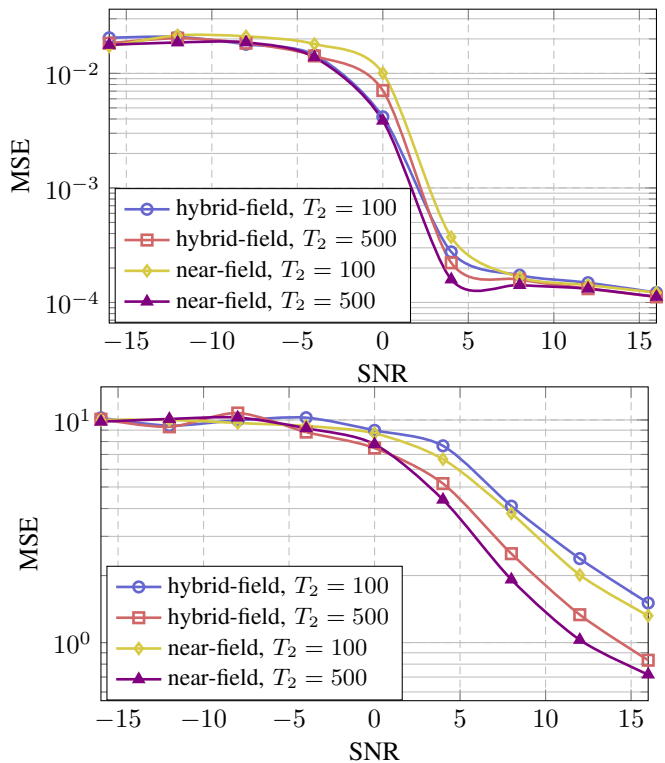

% \subsection{Channel Estimation} \label{sec:num_result_3}

Based on the possible angular sets obtained in \Cref{sec:num_result_2}, Fig.~\ref{fig:MSE_T_SNR} illustrates the variation of the angle MSE and distance MSE of the proposed refined algorithm with SNR and the number of samples $T_2$.
The number of sources is fixed at $K = 3$, each time block is divided into $S = 30$ slots, and the metasurface phase configuration is randomized in each sub-slot.
Two scenarios are evaluated: (i) a hybrid far-near-field case comprising two far-field and one near-field sources, and (ii) a pure near-field case where all three sources are in the near field.
The results show that the estimation MSE for both angles and distances decreases with increasing SNR, with the most pronounced improvement observed when the SNR exceeds $-5$~dB.
Notably, the angular estimation MSE for $T_2 = 100$ is nearly comparable to that for $T_2 = 500$, whereas the distance estimation MSE at $T_2 = 100$ still exhibits a visible performance gap relative to the $T_2 = 500$ case.
These observations verify the robustness of the proposed algorithm in both hybrid and pure near-field scenarios, as evidenced by the nearly overlapping MSE curves of the two cases.
\begin{figure}[t]
    \centering
    \hspace{-0.6cm} 
\begin{minipage}[t]{1\linewidth}
\centering
\begin{tikzpicture}        \renewcommand{\axisdefaulttryminticks}{4} 
    \pgfplotsset{every major grid/.style={densely dashed}}       
    \tikzstyle{every axis y label}+=[yshift=-5pt] 
    \tikzstyle{every axis x label}+=[yshift=-2pt]
    \pgfplotsset{every axis legend/.append style={cells={anchor=west},legend columns=2,fill=white, at={(-0.12,1.2)}, anchor=north west, font=\fontsize{7}{8.2}\selectfont}}
    \begin{axis}[
        width=0.95\columnwidth,
        height=0.75\columnwidth,
        ymajorgrids=true,
        scaled ticks=true,
        xlabel = {SNR},
        x label style={yshift=2pt},
        ylabel = {MSE (degree)},
        ylabel style={yshift=-1pt},
        ytick distance=2,
        ymode = log,
        ymin = 4*1e-6, ymax  = 1e+2,
        ytick={1e-6,1e-5,1e-4,1e-3,1e-2,1e-1,1e+0,1e+1,1e+2},
        grid=both,
        xmin  = -16,
        xmax = 16,
        x tick scale label style={
        font=\small,
        xshift=-5pt,
        yshift=-2pt    
        },
        ]
        \addplot[smooth, BLUE!60!white, line width=1pt, mark=diamond ] table[x=snr, y=mu-ang] {fig5_snr_near.txt}; 
        \addlegendentry{Angular-Modified MUSIC};
        \addplot[smooth, BLUE!60!white, line width=1pt, mark=square] table[x=snr, y=mu-dis] {fig5_snr_near.txt};
        \addlegendentry{Distance-Modified MUSIC};
        %-----
        \addplot[smooth, RED!60!white, line width=1pt, mark=diamond ] table[x=snr, y=es-ang] {fig5_snr_near.txt};
        \addlegendentry{Angular-the proposed algorithm};
        \addplot[smooth, RED!60!white, line width=1pt, mark=square] table[x=snr, y=es-dis] {fig5_snr_near.txt};
        \addlegendentry{Distance-the proposed algorithm};
        %-----
        \addplot[smooth,yellow!80!black, line width=1pt, mark=diamond ] table[x=snr, y=omp-ang] {fig5_snr_near.txt};
        \addlegendentry{Angular-HF OMP};
        \addplot[smooth,yellow!80!black, line width=1pt, mark=square] table[x=snr, y=omp-dis] {fig5_snr_near.txt};
        \addlegendentry{Distance-HF OMP};
        %-----
        \addplot[smooth, black, line width=1pt] table[x=snr, y=crb-ang] {fig5_snr_near.txt};
        \addlegendentry{Angular CRB};
        \addplot[smooth,gray, line width=1pt] table[x=snr, y=crb-dis] {fig5_snr_near.txt};
        \addlegendentry{Distance CRB};
    \end{axis}
\end{tikzpicture}
\end{minipage}%
\caption{ The MSE versus SNR.
The SNR is set to -5 dB. Results are averaged over 500 independent trials.
} 
\label{fig:vs}
\end{figure}
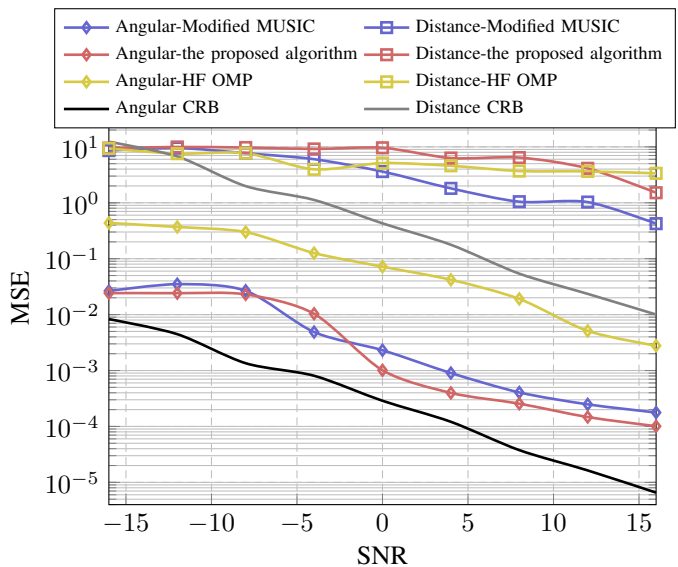

Finally, we compare the proposed channel estimation algorithm with two benchmark methods: the modified MUSIC algorithm introduced in \cite{he2011efficient} and the hybrid-field orthogonal matching pursuit (HF-OMP) approach proposed in \cite{wei2021channel}.
For the modified MUSIC implementation, spatial smoothing is performed by partitioning the metasurface into 144 subarrays, each with a dimension of $10 \times 10$. 
The simulation is conducted under the hybrid-field scenario with $K=3$ signals, $S=30$ slots and $T_2 = 100$ time blocks, while all other parameters are consistent with those in Fig.~\ref{fig:MSE_T_SNR}. 
The CRB curve is computed following the method in \cite{10681603}.
As shown in Fig.~\ref{fig:vs}, the proposed algorithm outperforms both benchmarks in terms of MSE. 
The performance of the HF-OMP algorithm is limited by the lack of orthogonality among atoms in its polar-domain dictionary, which degrades its resolution capability compared to the other super-resolution algorithms. 
The modified MUSIC algorithm, while effective, exhibits higher MSE compared to the proposed method because it employs a approximated near-field model that neglects second-order terms \cite{huang2023near}, a limitation that is explicitly accounted for in our proposed modeling framework.

\section{conclusion}
\label{sec:conclusion}

In this paper, we propose a low-complexity hardware solution for ELAA systems, a transceiver architecture integrating a minimal set of active antennas with a metasurface, referred to as MELA.
We derived a physically interpretable and mathematically tractable channel model for this architecture, establishing a solid foundation for performance analysis and algorithm design.
To enable efficient channel acquisition, a two-stage estimation scheme was developed: a coarse angular scanning stage to identify candidate directions, followed by a refinement stage that jointly estimates angles and ranges by exploiting subarray symmetry.
Furthermore, the MELA architecture achieves a HPBW comparable to that of conventional ELAA systems, while offering enhanced structural flexibility and reduced hardware weight.
This near-optimal resolution performance underscores the potential of MELA and motivates further investigation into digital precoding, which is not considered in this work.
Future research will focus on hybrid beamforming and transceiver optimization for MELA-based systems.

\begin{appendices}

\section{} \label{appd:A}
Let $\hat{\mathbf{u}}_{nk}^{(c)}\triangleq \frac{\mathbf{t}_n-\mathbf{p}_k}{\| \mathbf{t}_n-\mathbf{p}_k \|}-\frac{\mathbf{d}_c-\mathbf{t}_n}{\| \mathbf{d}_c-\mathbf{t}_n \|} $, and define $\mathbf{v}_{cn}=\mathbf{d}_c-\mathbf{t}_n$, $\hat{\mathbf{v}}_{cn}=\frac{\mathbf{d}_c-\mathbf{t}_n}{\| \mathbf{d}_c-\mathbf{t}_n \|}$.
The direction vector $\hat{\mathbf{v}}_{mn}$ can be approximated by its first-order expansion around $\hat{\mathbf{v}}_{cn}$ as
\begin{align*}
    \hat{\mathbf{v}}_{mn} \thickapprox \hat{\mathbf{v}}_{cn} + (\mathbf{I}-\hat{\mathbf{v}}_{cn}\hat{\mathbf{v}}_{cn}^\T) \bm{\delta}_m/\| \mathbf{v}_{cn}\|.
\end{align*}
Accordingly, the direction perturbation satisfies
\begin{align*}
    \| \Delta\hat{\mathbf{u}}_{nk} \| \triangleq  \hat{\mathbf{u}}_{mnk}-\hat{\mathbf{u}}_{nk}^{(c)} \thickapprox -(\mathbf{I}-\hat{\mathbf{v}}_{cn}\hat{\mathbf{v}}_{cn}^\T) \bm{\delta}_m/\| \mathbf{v}_{cn}\|,
\end{align*}
which yields the upper bound $\|\Delta\hat{\mathbf{u}}_{nk}\| \le D_1/\|\mathbf{v}_{cn}\|$.
The induced perturbation in $\mathcal{B}_{mnk}$ is proportional to the directional deviation, and can be bounded as
\begin{align*}
    \frac{|\Delta\mathcal{B}_{mnk} |}{|\mathcal{B}_{mnk}|} \lesssim C \frac{k_c d_t}{2} \frac{D_1}{\| \mathbf{v}_{cn} \|} \le \frac{2\pi D_1 D_h}{\lambda \| \mathbf{d}_c \|} \le \epsilon,
\end{align*}
where $C=\mathcal{O}(1)$ is a constant. 
Under the decoupled distance approximation, the right-hand side becomes negligible, confirming that replacing $\hat{\mathbf{v}}_{mn}$ with $\hat{\mathbf{v}}_{cn}$ introduces only a minimal modeling error in $\mathcal{B}_{mnk}$.
Consequently, $\mathcal{B}_{mnk}\approx \mathcal{B}_{nk}$ and is effectively independent of the receive index $m$.

\end{appendices}

\newpage

\bibliographystyle{IEEEtran}
\bibliography{ref}

\end{document}